\documentclass[12pt, draftclsnofoot, onecolumn]{IEEEtran}
\usepackage{mathdefs}
\usepackage{xspace}
\usepackage{graphicx}
\usepackage{amsmath}
\usepackage{amssymb}
\usepackage{amsthm}
\usepackage{hyperref}
\usepackage{kbordermatrix}
\usepackage{booktabs}
\usepackage{todonotes}
\usepackage{comment}

\newcommand{\ea}    {\emph{et al.}\xspace}

\newtheorem{corollary}{Corollary}

\newtheorem{lemma}{Lemma}
\newtheorem{proposition}{Proposition}

\newtheorem{theorem}{Theorem}
\newtheorem{remark}{Remark}

\newcommand{\corref}[1]{Cor.~\ref{cor:#1}}

\newcommand{\lemref}[1]{Lem.~\ref{lem:#1}}
\newcommand{\prpref}[1]{Prop.~\ref{prp:#1}}

\newcommand{\thmref}[1]{Thm.~\ref{thm:#1}}
\newcommand{\remref}[1]{Rem.~\ref{rem:#1}}

\renewcommand{\eqref}[1]{(\ref{eq:#1})}
\newcommand{\figref}[1]{Fig.~\ref{fig:#1}}
\newcommand{\tabref}[1]{Tab.~\ref{tab:#1}}

\newcommand{\secref}[1]{\S\ref{sec:#1}}
\newcommand{\prfref}[1]{App.~\ref{prf:#1}}

\DeclareMathOperator*{\argmin}{arg\,min}

\newcommand{\PPP}{\hat{\mathsf{\Phi}}}
\newcommand{\PPPb}{\tilde{\mathsf{\Phi}}}

\newcommand{\SINR}{\mathsf{\Sigma}}
\newcommand{\dT}{d_{\textup{T}}}
\newcommand{\rT}{r_{\textup{T}}}

\newcommand{\rO}{r_{\textup{O}}}
\newcommand{\expoA}{\lambda c_n \kappa_{\delta} \sigma^{\delta}}

\newcommand{\expoAN}{\sigma \eta}
\newcommand{\expoB}{\lambda c_n \rO^{n}}
\newcommand{\expoC}{\lambda c_n \sigma^{\delta} I(\chi,\delta)}
\newcommand{\pI}{p_{\textup{I}}}
\newcommand{\pII}{p_{\textup{II}}}

\newcommand{\rOMM}{r_{\textup{O,MM}}}
\newcommand{\rOEE}{r_{\textup{O,EE}}}
\newcommand{\rODI}{r_{\textup{O,DI}}}

\begin{document}

\title{On protocol and physical interference models in Poisson wireless networks}

\author{
Jeffrey~Wildman,~\IEEEmembership{Member,~IEEE}, 
Steven~Weber,~\IEEEmembership{Senior~Member,~IEEE}
\thanks{J. Wildman was with Drexel University while the majority of his contributions to this work was performed, and is now with MIT Lincoln Laboratory in Lexington, MA. S. Weber is with the Department of Electrical and Computer Engineering, Drexel University, Philadelphia, PA.  Support from the National Science Foundation (awards CNS-1147838 and CNS-1457306) is gratefully acknowledged. Preliminary versions of this work were presented at the Simons Conference on Networks and Stochastic Geometry (as a poster) in May, 2015 in Austin, TX \cite{WilWeb2015}, and at the International Symposium on Modeling and Optimization in Mobile, Ad Hoc and Wireless Networks (WiOpt) in May, 2016 in Tempe, AZ \cite{WilWeb2016}.  S.\ Weber is the contact author (\textsf{sweber@coe.drexel.edu}).}
}

\maketitle

\begin{abstract}
This paper analyzes the connection between the protocol and physical interference models in the setting of Poisson wireless networks. A transmission is successful under the protocol model if there are no interferers within a parameterized guard zone around the receiver, while a transmission is successful under the physical model if the signal to interference plus noise ratio (SINR) at the receiver is above a threshold.  The parameterized protocol model forms a family of decision rules for predicting the success or failure of the same transmission attempt under the physical model.  For Poisson wireless networks, we employ stochastic geometry to determine the prior, evidence, and posterior distributions associated with this estimation problem.  With this in hand, we proceed to develop five sets of results: $i)$ the maximum correlation of protocol and physical model success indicators, $ii)$ the minimum Bayes risk in estimating physical success from a protocol observation, $iii)$ the receiver operating characteristic (ROC) of false rejection (Type I) and false acceptance (Type II) probabilities, $iv)$ the impact of Rayleigh fading vs.\ no fading on the correlation and ROC, and $v)$ the impact of multiple prior protocol model observations in the setting of a wireless network with a fixed set of nodes in which the nodes employ the slotted Aloha protocol in each time slot.
\end{abstract}


\section{Introduction}
\label{sec:intro}

Interference models are a key component in the performance analysis of wireless networks due to the shared nature of the wireless medium. Several models have seen extensive use over the past several decades, including the \emph{physical} and \emph{protocol} interference models \cite{Car2010}. Successful reception under the physical interference model requires the signal to interference plus noise ratio (SINR) at the receiver exceed a threshold, while successful reception under the protocol interference model requires there be no interferers within a certain distance of the receiver.  

The key parameters in the physical and protocol models are the SINR threshold, denoted $\beta$, and the guard zone radius, denoted $\rO$.  Success (failure) under the physical model, i.e., receiver SINR above (below) $\beta$, is clearly distinct from success (failure) under the protocol model, i.e., interferers absent (present) from the disk of radius $\rO$ centered at the receiver.  Despite this distinction, the power law pathloss model for wireless transmission, i.e., $r^{-\alpha}$, suggests a positive correlation between these two events: low SINR is often due to high interference, which in turn is due to the presence of interferers near the receiver.  

With this in mind it is natural to seek to quantify the connection between the protocol and physical models in two ways: $i)$ the correlation between protocol and physical model success events, and $ii)$ the Bayes risk in predicting physical model success from protocol model observations.  The latter includes as a special case the receiver operating characteristic (ROC) between Type I (false rejection of) and Type II (false acceptance of) errors regarding the null hypothesis (physical model failure), given protocol model observations (the presence or absence of an interferer within $\rO$ of the receiver).  There is a tension in selecting $\rO$ to minimize the Bayes risk in this context: the presence (absence) of an interferer within a small $\rO$ gives strong (weak) evidence for physical model failure, while the absence (presence) of an interferer within a large $\rO$ gives strong (weak) evidence for physical model success.  We characterize $\rO$ that $i)$ maximizes the correlation of protocol and physical success, and $ii)$  minimizes the Bayes risk.  

\subsection{Related Work}
\label{sec:related-work}

Several works have explored how to employ the protocol model within the context of scheduling \cite{HasAnd2007, ShiHouLiu2013, ZhaCheLiu2014}. Hasan and Andrews \cite{HasAnd2007} study the protocol model as a scheduling algorithm in CDMA-based wireless ad hoc networks. They comment that a guard zone around each transmitter induces a natural tradeoff between interference and spatial reuse, affecting higher layer performance metrics such as transmission capacity, and they employ stochastic geometry to derive a guard zone that maximizes transmission capacity. Shi \ea~\cite{ShiHouLiu2013} examine the use of the protocol model within a cross-layer optimization framework and provide a strategy for correcting infeasible schedules generated under the protocol model by allowing transmission rate-adaptation to physical model SINR. Zhang \ea~\cite{ZhaCheLiu2014} analyze the effectiveness of protocol model scheduling using a variety of analytical, simulation, and testbed measurements.  This body of work on the protocol model as a scheduling paradigm is distinct from our focus on the protocol model as an {\em interference} model of the success or failure of attempted transmissions.  Iyer \ea~\cite{IyeRosKar2009} compares several interference models via simulation and qualitatively discusses the sacrifices in accuracy associated with abstracted interference models, including the protocol model. 

Finally, both the protocol and physical interference models have been studied within the framework of extremal and additive shot noise fields within stochastic geometry. Baccelli and B\l{}aszczyszyn \cite[Sec.~2.4]{BacBla2010} discuss the use of additive vs.\ extremal shot noise fields to model interference in wireless networks represented as point processes. Max (extremal) interference finds use in bounding outage and transmission capacity under sum (additive) interference in PPPs with Aloha scheduling, as is done in \cite[Sec.~2.5]{WebAnd2012}.

\subsection{Contributions and outline}

The outline of the paper is as follows. \secref{model} introduces the model and notation; \secref{prevpo} derives the prior, evidence, and posterior distributions for protocol and physical model success; \secref{corr} derives the correlation of the protocol and physical success events; \secref{bin-hypo-test} derives the Bayes risk in predicting physical success from protocol observations; \secref{roc} specializes the Bayes risk to the uniform cost model, with the ROC parameterized by $\rO$; \secref{fading} studies the impact of Rayleigh fading on the correlation and ROC by contrasting with the case of no fading; \secref{multiobs} addresses the case when a fixed set of potential transmitters employ the slotted Aloha protocol, and studies the impact of multiple prior protocol model observations on the optimal prediction of physical model success; finally \secref{conclusions} holds a brief conclusion.  Longer proofs are in the Appendix.

The primary contributions are as follows.  \prpref{coroptobs} (\secref{corr}) characterizes the $\rO$ to maximize the correlation of protocol and physical model success events; \thmref{min-risk} (\secref{bin-hypo-test}) characterizes the $\rO$ to minimize the Bayes risk in predicting physical model success from protocol model observations; \prpref{errorprob} (\secref{roc}) specializes the Bayes risk model to obtain the ROC of Type I vs.\ Type II errors; \prpref{nofadelikelihood} (\secref{fading}) gives a numerical means of computing the ROC for the case of no fading, and \figref{nofading} demonstrates the impact of fading can be significant; finally \thmref{physucconmindisk}, \prpref{multipriorprot}, \prpref{multievidence} (\secref{multiobs}) enable computation of the ROC under multiple prior protocol model observations, and \figref{multipleobs} suggests these observations may be ignored under the optimal decision rule.

\section{Model}
\label{sec:model}

Random variables (RVs) are given a sans-serif font, e.g., $\xsf,\msf$.  We use the standard acronyms for independent and identically distributed (IID), probability density / mass function (PMF/PDF), cumulative distribution function (CDF), complementary CDF (CCDF), Laplace transform (LT), and inverse LT (ILT).  Probability is written $\Pbb(\cdot)$, expectation is written $\Ebb[\cdot]$, and the LT of $\xsf$ with PDF $f$ is written $\Lmc_{\xsf}(s) = \Lmc[f](s)$.  A bar denotes complement: $\bar{p}(\cdot) \equiv 1 - p(\cdot)$.  

Transmitter and receiver are abbreviated as TX and RX, respectively.  Euclidean distance of a point $x \in \Rbb^n$ from the origin is denoted $\|x\|$, and the ball in $\Rbb^n$ of radius $r$ is denoted $b(o,r)$. Natural and real numbers are denoted by $\Nbb$ and $\Rbb$, respectively.  All logs are natural.  Denote $\{1,\ldots,N\}$ by $[N]$, for $N \in \Nbb$.  The indicator $\mathbf{1}_A$, for any statement $A$, equals $1$ ($0$) if $A$ is true (false).  The notation $A \equiv B$ means $A=B$ by definition.  \tabref{notation} lists notation.

\begin{table}
\centering
\caption{Notation}
\begin{tabular}{ll} \toprule
Symbol & Meaning \\ \midrule
$n$ & ambient dimension ($n \in \{1,2,3\}$) \\
$\PPP = \{(\xsf_i,\msf_i)\}$ & homogeneous marked PPP of TX, RX locations \\
$\msf_i = (\zsf_i,\Fsf_i)$ & mark for TX-RX pair $i$ \\
$\zsf_i$ & location of RX $i$ relative to TX $i$ \\
$\Fsf_i$ & Rayleigh fade from TX $i$ to reference RX \\

$(\xsf_i,\ysf_i)$ & TX $i$ at $\xsf_i$ and RX $i$ at $\ysf_i = \xsf_i + \zsf_i$ \\
$\lambda$ & (spatial) intensity of $\PPP$ \\
$\rT$ & TX-RX separation distance \\
$(\xsf_o,\ysf_o)$ & location of reference TX and RX \\
$\Psf_i$ & received power from TX $i$ at reference RX \\
$\alpha$ & large-scale pathloss constant \\
$\beta$ & SINR threshold \\
$\SINR_o$ & SINR at reference RX \\
$\Isf_o$ & sum interference at reference RX \\
$\eta$ & background noise power \\
$\Hsf$ & RV for reference transmission physical model success/failure \\
$\rO$ & protocol model guard zone / observation radius \\
$\Dsf$ & RV for reference transmission protocol model success/failure \\
$c_n$ & volume of a unit ball in $\Rbb^n$ \\
$\delta$ & characteristic exponent $n/\alpha$ \\
$\kappa_{\delta}$ & convenience parameter \eqref{kappadef} \\
$\sigma$ & convenience parameter $\beta\rT^\alpha$ \\
$\chi$ & convenience parameter $\rO^{\alpha}/\sigma = (\rO/\rT)^{\alpha}/\beta$ \\
$I(u,\delta)$ & convenience function \eqref{convfcn} \\ 
$g$ & decision rule $g$ maps observations $\Dsf$ to predictions of $\Hsf$ \\
$(g,\rO)$ & decision rule pair, the control parameters \\
$\cbf$ & cost matrix for Bayesian risks \eqref{costmatrixdef} \\
$R_i(g,\rO)$ & conditional risks ($i \in \{1,2\}$) of decision rule $(g,\rO)$ \eqref{condriskdef} \\
$R(g,\rO)$ & Bayes risk (expected cost) of decision rule $(g,\rO)$ \eqref{risk2} \\
$\rO^*(g)$ & Bayes-optimal radius for rule $g$ \eqref{min-bayes-risk-problem} \\
$A,B,C$ & variables used to express the Bayes risk $R(g,\rO)$ \eqref{ABC} \\
$Q(z)$ & CCDF of a standard normal $\Zsf \sim \Nmc(0,1)$ \\ 
$\mu_d$ & convenience parameter $\lambda c_n \rO^n$ \\
$\xi$ & convenience parameter $\frac{p}{\bar{p}} \chi^{-\delta} I(\chi,\delta)$ \\\bottomrule
\end{tabular}
\label{tab:notation}
\end{table}

\subsection{Poisson model of instantaneous node locations}

Let $n \in \{1,2,3\}$ denote the ambient dimension of the network.  We model the instantaneous locations of the nodes comprising the wireless network by the marked, bipolar, homogeneous Poisson Point Process (PPP) $\PPP = \{(\xsf_i,\msf_i), i \in \Nbb\}$ in $\Rbb^n$ of intensity $\lambda > 0$.  The term {\em bipolar} means we assume a pairing / matching of transmitters with receivers.  The point $\xsf_i$ and mark $\msf_i$, with $\msf_i \equiv (\zsf_i,\Fsf_i)$, correspond to the $i^{\rm th}$ TX-RX pair, with the TX at location $\xsf_i$ and the RX at location $\ysf_i = \xsf_i + \zsf_i$.  The TX locations $\{\xsf_i\}$ form a homogeneous PPP of intensity $\lambda$, and the mark components $\{\zsf_i\}$ are IID on the $n$-dimensional sphere with TX-RX separation distance $\rT$.  We will require the transmission success probability of a reference TX-RX pair at $(\xsf_o,\ysf_o)$, with the reference RX at the origin $o$.  Slivnyak's Theorem \cite[Thm.~8.1]{Hae2013}, applied to the PPP $\PPP$, ensures the reduced Palm distribution of $\PPP$ is equal in distribution to the original $\PPP$.

\subsection{Physical interference model}
\label{sec:model-physical}

We assume a (standard) signal propagation model for large-scale, distance-based pathloss with Rayleigh fading, and unit transmission power. The signal power at RX $o$ from TX $i$ is $\Psf_i \equiv \Fsf_i l(\|\xsf_i\|)$, for $\Fsf_i \sim \text{Exp}(1)$ the (random) Rayleigh fading from TX $i$ to RX $o$, $l(r) \equiv r^{-\alpha}$ the large-scale pathloss function with pathloss exponent $\alpha > n$, and $\|\xsf_i\|$ the (random) distance from TX $i$ to RX $o$ (note: $\|\xsf_o\| = \rT$, by assumption).  The fading RVs $\{\Fsf_i\}$ are IID.  A transmission between the reference TX-RX pair $o$ is considered successful under the physical interference model if the (random) SINR at the reference RX, denoted $\SINR_o$, exceeds an SINR threshold $\beta > 0$, with $\SINR_o \equiv \Psf_o/(\Isf_o + \eta)$, $\eta \geq 0$ the background noise power, and $\Isf_o \equiv \sum_{i \neq o} \Psf_i$ the (random) sum interference power at RX $o$. The Bernoulli RV $\Hsf \equiv \mathbf{1}\{\SINR_o \geq \beta\}$ represents physical model success or failure of the reference transmission $o$ in $\PPP$.  The corresponding events (hypotheses) that the reference transmission fails (succeeds) under the physical model are $\{ \Hsf = 0 \}$ and $\{ \Hsf = 1 \}$.

\subsection{Protocol interference model}
\label{sec:model-protocol}

We also employ a (standard) protocol interference model, characterized by a guard zone distance\footnote{Alternately, one may employ a guard zone factor $\Delta$ of the TX-RX distance $\rT$, producing (potentially unique) guard zone distances: $\rO = (1+\Delta)\rT$. Under our model with a fixed TX-RX $\rT$, these formulations are equivalent.} $\rO$. A transmission between TX-RX pair $o$ is considered successful under the protocol interference model iff there are no interfering TX's within distance $\rO$ of the reference RX at $o$.  The Bernoulli RV $\Dsf = \Dsf(\rO) \equiv \mathbf{1}\{\|\xsf_i\| \geq \rO, \forall i \neq o\}$ represents the success or failure under the protocol model of the reference transmission $o$ in $\PPP$, with corresponding events (observations) that the reference transmission fails (succeeds) under the protocol model: $\{ \Dsf(\rO) = 0 \}$ and $\{ \Dsf(\rO) = 1 \}$.
We treat $\rO$ as a control parameter on the observation $\Dsf$, as described in \secref{bin-hypo-test}. 

\subsection{Special functions and convenience parameters}
\label{sec:model-notation}
We use the Gamma and generalized exponential functions:
\begin{equation}
\Gamma(v) \equiv \int_{0}^{\infty} t^{v-1} \erm^{-t} \drm t, ~ 
E(v,u) \equiv \int_{1}^{\infty} \erm^{-ut} t^{-v} \drm t \label{eq:stdfcndef}
\end{equation}
Define notation: $i)$ $c_n$ is the volume of a unit ball in $\Rbb^n$ ($c_1 = 2$, $c_2 = \pi$, and $c_3 = 4 \pi/3$), $ii)$ $\delta \equiv n/\alpha$ is the characteristic exponent ($\delta < 1$ is assumed), $iii)$ the convenience function
\begin{equation}
\label{eq:kappadef}
\kappa_{\delta} \equiv \Gamma(1+\delta)\Gamma(1-\delta) = \frac{\pi \delta}{\sin(\pi \delta)} = \delta \int_0^{\infty} \frac{t^{\delta-1}}{1+t}\drm t
\end{equation}
is convex increasing in $\delta$ over $[0,1)$ with $\kappa_0 = 1$ and $\lim_{\delta \uparrow \infty} \kappa_{\delta} = \infty$, $iv)$ $\sigma \equiv \beta\rT^\alpha$ is a convenience parameter, $v)$ $\chi \equiv  \rO^{\alpha}/\sigma = (\rO/\rT)^{\alpha}/\beta$ is a convenience parameter, and $vi)$ 
\begin{equation}
\label{eq:convfcn}
I(u,\delta) \equiv 
\delta \int_{0}^{u} \frac{t^\delta}{1+t} \drm t
\end{equation}
obeys $I(0,\delta) = 0$, $\frac{\drm}{\drm u} I(u,\delta) = \delta \frac{u^{\delta}}{1+u} \geq 0$, and $\lim_{u \uparrow \infty} I(u,\delta) = \infty$.\footnote{For computation it is useful to note that $I(u,\delta) = (-1)^{1-\delta} \delta B(-u,1+\delta,0)$, for $B$ the incomplete beta function.}

\section{The prior, evidence, and posterior distributions}
\label{sec:prevpo}

Having introduced definitions and notation, we now provide several results.  \lemref{prior} gives the prior distribution on $\Hsf$, $p_{\Hsf}(h)$, \lemref{evidence} gives the evidence distribution on $\Dsf$, $p_{\Dsf}(d) \equiv \Pbb(\Dsf = d)$, and \prpref{posterior} gives the posterior distribution of $\Hsf$ given $\Dsf$, $p_{\Hsf|\Dsf}(h|d) \equiv \Pbb(\Hsf=h|\Dsf=d)$. 

\begin{lemma}
\label{lem:prior}
The (prior) distribution of the physical model feasibility RV $\Hsf$ is:
\begin{equation}
p_{\Hsf}(1) = \exp(-\expoA -\expoAN).
\end{equation}
\end{lemma}

\begin{IEEEproof}
This result follows from standard stochastic geometry arguments on the outage probability of the power law pathloss function with Rayleigh fading for a PPP \cite[p. 104]{Hae2013}; it may also be obtained by letting the void zone radius approach zero ($\rO \downarrow 0$) in \corref{laplace-with-void}.
\end{IEEEproof}

\begin{lemma}
\label{lem:evidence}
The (evidence) distribution of the protocol model feasibility RV $\Dsf$ is:
\begin{equation}
p_{\Dsf}(1) = \exp(-\expoB).
\end{equation}
\end{lemma}

\begin{IEEEproof}
$p_{\Dsf}(1)$ is the void probability of the guard zone of radius $\rO$ \cite[Thm.~2.24]{Hae2013}.
\end{IEEEproof}

\begin{proposition}
\label{prp:posterior}
The (posterior) distribution of $\Hsf$ given $\Dsf$ is:
\begin{equation}
\label{eq:post-h1-given-d1}
p_{\Hsf|\Dsf}(1|1) = \erm^{-A+B(\rO)-C(\rO)}
\end{equation}
where (with $\chi \equiv \rO^{\alpha}/\sigma$): $A \equiv \expoA + \expoAN$, $B(\rO) \equiv \expoB$, and $C(\rO) \equiv \expoC$.
\end{proposition}

The proof is in \prfref{posterior}.  The quantities $p_{\Hsf|\Dsf}(1|0),p_{\Hsf|\Dsf}(0|1),p_{\Hsf|\Dsf}(0|0)$ are expressible in terms of $p_{\Hsf|\Dsf}(1|1), p_{\Hsf}(1),p_{\Dsf}(1)$. For example: $p_{\Hsf|\Dsf}(1|0) = (p_{\Hsf}(1) - p_{\Hsf|\Dsf}(1|1)p_{\Dsf}(1))/\bar{p}_{\Dsf}(1)$.  Moreover, $A$, $B(\rO)$, $C(\rO)$ in \eqref{post-h1-given-d1} obey: $p_{\Hsf}(1) = \erm^{-A}$, $p_{\Dsf}(1) = \erm^{-B(\rO)}$, and $p_{\Hsf|\Dsf}(1|1)p_{\Dsf}(1) = \erm^{-A-C(\rO)}$.

The posterior distribution is not well-defined for $\rO = \{0,\infty\}$: when $\rO \downarrow 0$ ($\rO \uparrow \infty$) the event $\Dsf=0$ ($\Dsf=1$) occurs with probability $0$.  Neither case affects our analysis.  

\section{Correlation of $\Hsf,\Dsf$}
\label{sec:corr}

We leverage \lemref{prior}, \lemref{evidence}, and \prpref{posterior} to compute the correlation of $(\Hsf,\Dsf)$, denoted $\rho = \rho_{\Hsf,\Dsf}$.  For the following result, the proof of which is found in \prfref{coroptobs}, it is convenient to use the change of variable from $\rO$ to $\chi = \chi(\rO) \equiv \rO^{\alpha}/\sigma$.  With this change, $B(\rO),C(\rO)$ in \prpref{posterior} become $B(\chi) \equiv \lambda c_n \sigma^{\delta} \chi^{\delta}$, $C(\chi) \equiv \lambda c_n \sigma^{\delta} I(\chi,\delta)$.  Applying the definition of correlation to Bernoulli RVs and substituting the above results gives
\begin{eqnarray}
\rho_{\Hsf,\Dsf}(\chi)
&\equiv& \frac{\Ebb[\Hsf \Dsf]-\Ebb[\Hsf]\Ebb[\Dsf]}{\sqrt{\mathrm{Var}(\Hsf)\mathrm{Var}(\Dsf)}} \nonumber \\
&=& \left(\frac{p_{\Hsf|\Dsf}(1|1)}{p_{\Hsf}(1)}-1\right) \sqrt{\frac{p_{\Hsf}(1) p_{\Dsf}(1)}{\bar{p}_{\Hsf}(1) \bar{p}_{\Dsf}(1)}} \nonumber \\
&=& \frac{\erm^{B(\chi)-C(\chi)}-1}{\sqrt{(\erm^A-1)(\erm^{B(\chi)}-1)}}. \label{eq:corHDexp}
\end{eqnarray}

\begin{proposition}
\label{prp:coroptobs}
The correlation $\rho(\chi)$ obeys:
$i)$ $\lim_{\chi \downarrow 0} \rho(\chi) = 0$;
$ii)$ $\lim_{\chi \uparrow \infty} \rho(\chi) = 0$;
$iii)$ $\rho(\chi) \in (0,1]$ for all $\chi > 0$;
$iv)$ has a maximum at $\chi^* > 1$ equal to a positive solution of
\begin{equation}
\label{eq:coroptobs}
(1-\chi) \erm^{B(\chi)} + (1+\chi)\erm^{C(\chi)} = 2.
\end{equation}
\end{proposition}

Numerical experiments suggest the following are true: $i)$ $\rho(\chi)$ has a single stationary point, i.e., $\eqref{coroptobs}$ has a unique solution for $\chi > 0$, and $ii)$ there is a unique inflection point $\chi^{**} > \chi^*$ solving $\rho^{''}(\chi) = 0$.  These ensure $iii)$ $\rho(\chi)$ is concave increasing in $\chi$ over $[0,\chi^*]$, $iv)$ concave decreasing in $\chi$ over $[\chi^*,\chi^{**}]$, and $v)$ convex decreasing in $\chi$ over $[\chi^{**},\infty)$.  

\begin{figure}[ht]
\centering
\includegraphics[width=0.75\linewidth]{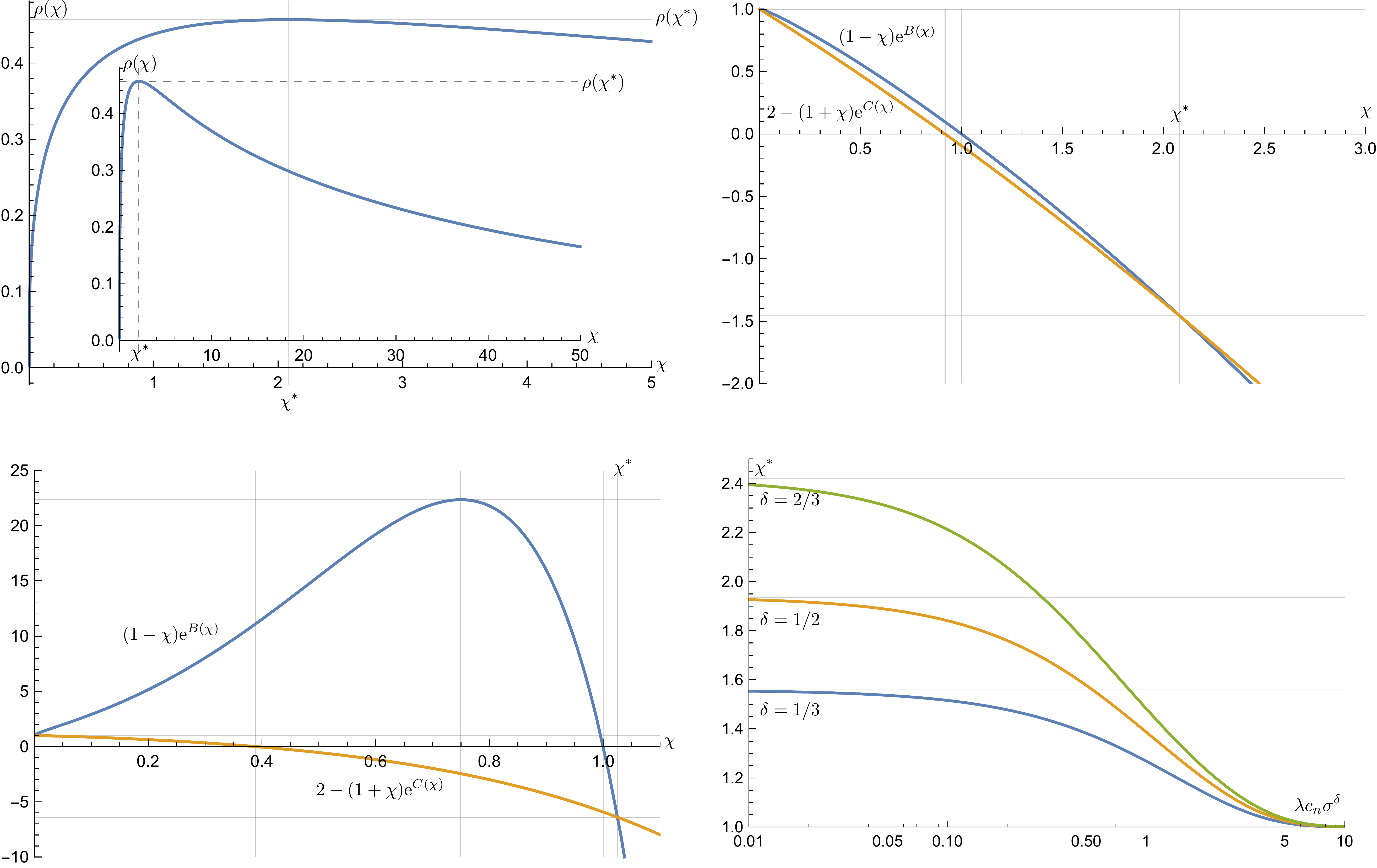}
\caption{
{\bf Top left:} correlation of $(\Hsf,\Dsf)$, $\rho_{\Hsf,\Dsf}(\chi)$, vs. $\chi \equiv \rO^{\alpha}/\sigma$, for $n=2, \lambda = 2 \times 10^{-4}, \alpha = 3, \beta = 5, \rT = 10, \eta = 0$: the maximum correlation is at $\chi^* \approx 2.08$ and the inset suggests that, for large $\chi$, $\rho(\chi)$ is convex decreasing in $\chi$, with $\lim_{\chi \uparrow \infty} \rho(\chi) = 0$.  
{\bf Top right:} the function $f_1(\chi) \equiv (1-\chi) \erm^{B(\chi)}$ and the concave decreasing function $f_2(\chi) \equiv 2 - (1+\chi) \erm^{C(\chi)}$ vs.\ $\chi$ for the same values, where $\chi^*$ is the unique positive value such that $f_1(\chi^*) = f_2(\chi^*)$.  
{\bf Bottom left:} $f_1(\chi),f_2(\chi)$ for same values but with $\lambda$ replaced with $1/\lambda$: $f_1(\chi)$ has a very small concave neighborhood near zero (not visible), followed by a convex neighborhood, and is thereafter concave.
{\bf Bottom right:} the optimal $\chi^*$ as a function of $\lambda c_n \sigma^{\delta}$ for $\delta \in \{1/3,1/2,2/3\}$; 
}
\label{fig:correlationHD}
\end{figure}

\figref{correlationHD} illustrates the functions $\rho(\chi)$, $f_1(\chi) \equiv (1-\chi) \erm^{B(\chi)}$, and (concave decreasing) $f_2(\chi) \equiv 2 - (1+\chi) \erm^{C(\chi)}$ vs.\ $\chi$  (observe \eqref{coroptobs} is equivalent to $f_1(\chi) = f_2(\chi)$).  The top two plots are for ``typical parameters'', while the bottom left, while admittedly atypical, illustrate some of the structure of $f_1(\chi)$ not visible in the previous case.  Finally, the bottom right plot shows the optimized $\chi^*$ as a function of $\lambda c_n \sigma^{\delta}$.  The gridlines showing $\chi^*$ as $\lambda c_n \sigma^{\delta} \downarrow 0$ are easily seen to be the solution of $I(\chi,\delta) = \frac{\chi-1}{\chi+1}\chi^{\delta}$.

\section{Bayes risk for binary hypothesis testing}
\label{sec:bin-hypo-test}

We employ a Bayesian binary hypothesis testing framework, where the two hypotheses $\Hsf$  represent the possible ``ground truth'' under the physical model, and $\Dsf$ represents the two possible observations under the protocol model.  A decision rule $g(d) : \{0,1\} \to \{0,1\}$ in this case maps observations $\Dsf = d$ to predictions $\Hsf = h$.  Let $\Gmc$ ($|\Gmc|=4$) be the set of rules $g$, each of which is parameterized by $\rO$, in the following manner: {\em observe $\Dsf(\rO) = d$ then predict $\Hsf = h = g(d)$}. Thus, $g$ predicts the corresponding physical model outcome, $\Hsf$, given the observed protocol model outcome, $\Dsf(\rO)$. The pair $(g,\rO)$ are control parameters, and the suitability of a rule $g \in \Gmc$ and a radius $\rO$ as a predictor for $\Hsf$ will vary with $(g,\rO)$.

Define a nonnegative cost matrix 
\begin{equation}
\label{eq:costmatrixdef}
\cbf \equiv \kbordermatrix{
     & \Hsf = 0 & \Hsf = 1 \\
g(\Dsf) = 0 & c_{00} & c_{01}  \\
g(\Dsf) = 1 & c_{10} & c_{11}
},
\end{equation}
where $c_{ij} \geq 0$ is the cost of making decision $g(\Dsf) = i$ when hypothesis $\Hsf = j$ is true.  In \secref{roc} we will specialize to the {\em uniform cost model} that does not penalize correct decisions and uniformly penalizes incorrect decisions: $c_{01} = c_{10} = 1$ and $c_{00} = c_{11} = 0$.

Using the cost matrix, we may enumerate the conditional risks, $R_0(g,\rO)$ and $R_1(g,\rO)$, associated with the decision rule $(g,\rO)$ with notation $p_{g(\Dsf)|\Hsf}(h'|h) \equiv \Pbb(g(\Dsf)=h'|\Hsf=h)$:
\begin{eqnarray}
R_0(g,\rO) & \equiv & c_{10} p_{g(\Dsf)|\Hsf}(1|0) + c_{00} p_{g(\Dsf)|\Hsf}(0|0) \nonumber \\
R_1(g,\rO) & \equiv & c_{11} p_{g(\Dsf)|\Hsf}(1|1) + c_{01} p_{g(\Dsf)|\Hsf}(0|1). \label{eq:condriskdef}
\end{eqnarray}
These risks provide the expected costs of decision rule $(g,\rO)$ conditioned on the value of $\Hsf$, the RV to be estimated.  Under the uniform cost model (\secref{roc}), $R_0(g,\rO)$ and $R_1(g,\rO)$ yield the false rejection (Type I error) rate and the false acceptance (Type II error) rate, respectively.

The total expected cost, i.e., Bayes risk, of decision rule $(g,\rO)$, is (with $p_{\Hsf}(h) \equiv \Pbb(\Hsf = h)$, $p_{g(\Dsf)}(h') \equiv \Pbb(g(\Dsf) = h')$, and $p_{\Hsf|g(\Dsf)}(h|h') \equiv \Pbb(\Hsf = h|g(\Dsf)=h')$:
\begin{eqnarray}
\label{eq:risk2}
R(g,\rO) &\equiv& R_0(g,\rO) \bar{p}_{\Hsf}(1) + R_1(g,\rO) p_{\Hsf}(1) \label{eq:risk} \nonumber \\
&=& c_{00} + (c_{01} - c_{00})p_{\Hsf}(1)  + (c_{10} - c_{00}) p_{g(\Dsf)}(1) \nonumber \\
& & + (c_{11} + c_{00} - c_{01} - c_{10}) p_{\Hsf|g(\Dsf)}(1|1)p_{g(\Dsf)}(1). 
\end{eqnarray}
A Bayes-optimal observation radius $\rO^*(g)$ for rule $g$ minimizes the incurred Bayes risk:
\begin{equation}
\label{eq:min-bayes-risk-problem}
\rO^*(g) \in \argmin_{\rO \geq 0} R(g,\rO). 
\end{equation}
\prpref{risk} gives the Bayes risk $R(g,\rO)$ in terms of the model parameters, and our first main result, \thmref{min-risk}, gives the Bayes-optimal decision rule parameter $\rO^*(g)$.

For the remainder of this section we restrict the decision rule to the identity map $g(d) = d$, meaning observed protocol model success (failure) predicts physical model success (failure).  As $|\Gmc|=4$, discounting the two constant rules ($g(d) = 0$ and $g(d)=1$), and recalling $\rho_{\Hsf,\Dsf} > 0$, the identity map is clearly superior to the only other decision rule, the complement map $g(d) = \bar{d}$.

\begin{proposition}
\label{prp:risk}
Under the identity decision rule $g(d) = d$, the Bayes risk $R(\rO)$ \eqref{risk2} of decision rule parameter $\rO$ is (with $A,B,C$ in \prpref{posterior}):
\begin{eqnarray}
R(\rO) &=& c_{00} + (c_{01} - c_{00})\erm^{-A} + (c_{10} - c_{00})\erm^{-B(\rO)} \nonumber \\
& & + (c_{11} + c_{00} - c_{10} - c_{01})\erm^{-A-C(\rO)}. \label{eq:bayes-risk-ppp}
\end{eqnarray}
\end{proposition}

\begin{IEEEproof}
The result follows from \eqref{risk2} and expressions from \lemref{prior}, \lemref{evidence}, and \prpref{posterior}.
\end{IEEEproof}

\begin{theorem}
\label{thm:min-risk}
Under the identity decision rule $g(d) = d$, if $c_{10} > c_{00}$ and $c_{01} > c_{11}$, the optimal radius $\rO^*$ \eqref{min-bayes-risk-problem} minimizing the  risk $R(\rO)$ \eqref{risk2} is the unique solution to (with $A,B,C$ in \prpref{posterior}):
\begin{equation}
\label{eq:gz-min-risk}
\frac{1}{1 + \frac{c_{01}-c_{11}}{c_{10}-c_{00}}} \left(1+\frac{1}{\chi(\rO)}\right) = \exp(-A + B(\rO) - C(\rO)).
\end{equation}
This solution exists iff 
\begin{equation}
\label{eq:existtest}
\log\left(1+\frac{c_{01}-c_{11}}{c_{10}-c_{00}}\right) > \sigma \eta.
\end{equation}
The risk $R(\rO)$ is quasi-convex but not convex in $\rO$.  The minimized risk is:
\begin{equation}
\label{eq:optrisk}
R(\rO^*) = c_{00} + (c_{01} - c_{00}) \erm^{-A} - (c_{10} - c_{00}) \frac{1}{\chi(\rO^*)} \erm^{-B(\rO^*)}.
\end{equation}
\end{theorem}

The proof is in \prfref{min-risk}.  The conditions $c_{10} > c_{00}$ and $c_{01} > c_{11}$ mean the cost of a wrong decision exceeds the cost of a correct decision, a typical assumption in a Bayes estimation framework. The coefficient ratio on the left side of \eqref{gz-min-risk} equals $1/2$ under the uniform cost model.  \prpref{sensitivity} gives the change in $\rO^*$ with respect to changes in $(\lambda,\sigma)$.  As $\sigma \equiv \rT^{\alpha}/\beta$, the sensitivity with respect to both $(\rT,\beta)$ is easily obtained from the sensitivity with respect to $\sigma$.

\begin{proposition}
\label{prp:sensitivity}
Under the identity decision rule $g(d) = d$, the sensitivities of $\rO^*$ to changes in $(\lambda,\sigma)$ are (assuming $\eta = 0$):
\begin{eqnarray}
\frac{\drm \rO^* }{\drm \lambda} &=& \frac{c_n \rO (1+\chi)(\kappa_{\delta} + I(\chi,\delta) - \chi^{\delta})}{\alpha (1 + c_n \delta \lambda \rO^n)}  \label{eq:dIstdlam} \\
\frac{\drm \rO^*}{\drm \sigma}  &=& \frac{\rO \left(1 + \lambda c_n \delta \sigma^{\delta} \left[ (1+\chi)(\kappa_{\delta} + I(\chi,\delta)) - \chi^{\delta+1} \right] \right)}{\alpha \sigma(1 + c_n \delta \lambda \rO^n)}  \label{eq:dIstds}
\end{eqnarray}
for $\chi \equiv \rO^{\alpha}/\sigma$.  Moreover, $\frac{\drm \rO^*}{\drm \lambda} > 0$ and $\frac{\drm \rO^*}{\drm \sigma} > 0$.
\end{proposition}

\begin{figure}
\centering
\includegraphics[width=0.75\linewidth,keepaspectratio]{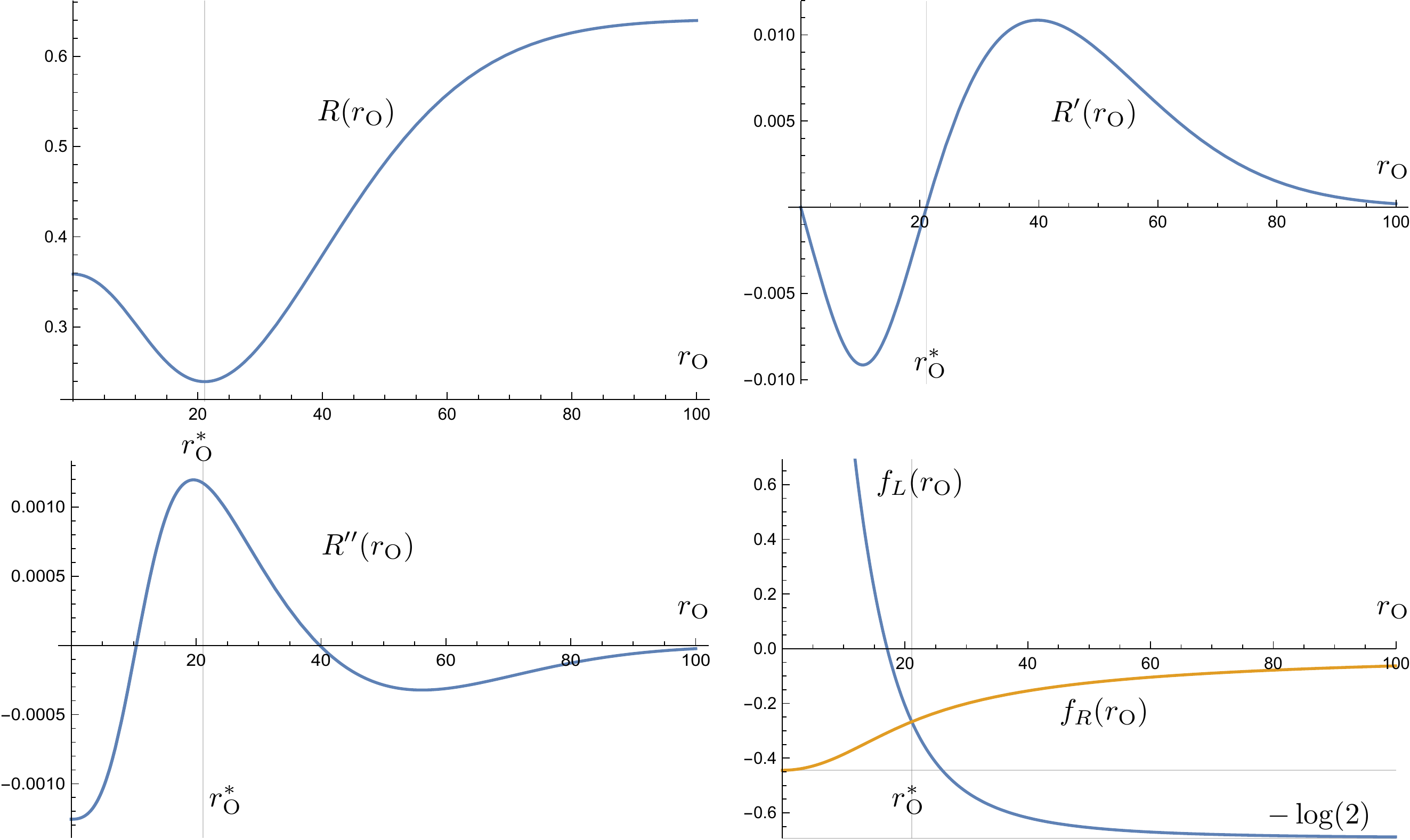}
\caption{The Bayes risk $R(\rO)$ (top left), its first two derivatives, $R'(\rO)$ (top right) and $R''(\rO)$ (bottom left), and the functions $f_L(\rO),f_R(\rO)$ (bottom right), all vs.\ $\rO$. Parameters: $n=2$, $\lambda = 2 \times 10^{-4}$, $\alpha = 3$, $\dT = 10$, $\beta = 5$,  $\eta = 0$, and $\cbf$ for the uniform cost model (yielding $\gamma = \nu = 1$ in the proof of \thmref{min-risk}).  $R(\rO)$ is quasi-convex, but not convex, with unique minimizer $\rO^*$ the solution of $R'(\rO) = 0$ (equivalently, $f_L(\rO) = f_R(\rO)$).}
\label{fig:BayesRisk}
\end{figure}

\begin{figure}
\centering
\includegraphics[width=0.75\linewidth,keepaspectratio]{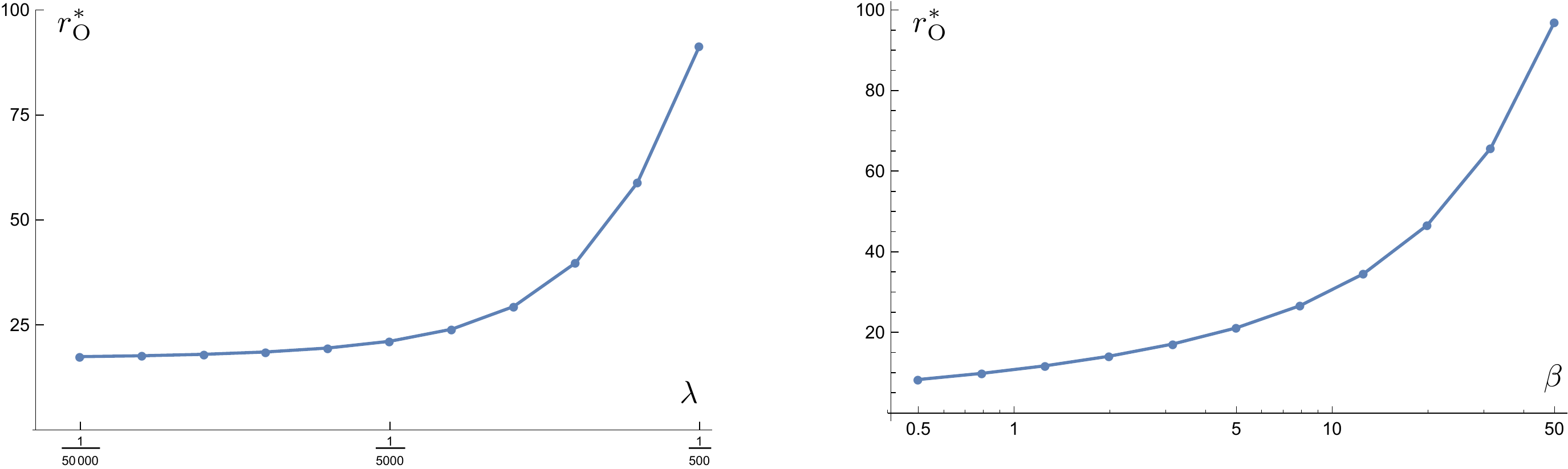}
\caption{The optimized radius $\rO^*$ vs. $\lambda$ (left) and $\beta$ (right).  As shown in \prpref{sensitivity}, $\frac{\drm}{\drm \lambda} \rO^* > 0$ and $\frac{\drm}{\drm {\sigma}} \rO^* > 0$ (recall ${\sigma} \equiv \beta \rT^{\alpha}$).}
\label{fig:OptIDSens}
\end{figure}

The proof is in \prfref{sensitivity}.  The restriction to the no-noise case $\eta = 0$ is only to slightly simplify the resulting expressions; the sensitivities for general $\eta$ may be easily derived.  The Bayes risk expressions are illustrated in \figref{BayesRisk}, and the sensitivities of $\rO^*$ to $\lambda,\beta$ are shown in \figref{OptIDSens}.

\begin{remark}
Both $\frac{\drm }{\drm \lambda} \rO^* > 0$ and $\frac{\drm }{\drm \beta} \rO^* > 0$ are intuitive, but the former is not obvious.  First, the probability of a physical model failure increases with $\lambda$ (due to increased interference); as such the protocol model failure probability must likewise be increased in order to minimize the overall risk.  Increasing $\rO$ achieves this goal, but observe the protocol failure probability is {\em already} increasing in $\lambda$ for {\em fixed} $\rO$, since the void probability of the observation disk is decreasing.  Apparently, it must be additionally increased by expanding $\rO$.  Second, the probability of a physical model failure increases with $\beta$ (due to the higher required SINR), and as such, again, the protocol model failure probability must be increased, which explains why $\rO^*$ increases.
\end{remark}

\section{Uniform cost model ROC}
\label{sec:roc}

Given \thmref{min-risk}, we know how to find the guard zone $\rO^*$ that minimizes the protocol model's Bayes risk associated with predicting physical model feasibility. Deviating from $\rO^*$ in either direction will result in an increase in the average risk $R(\rO)$, but will also trade off the two types of conditional risk, $R_0(\rO)$ and $R_1(\rO)$. We analyze this tradeoff under the uniform cost model (the receiver operating characteristic (ROC)) for each decision rule $g \in \Gmc$.  The ROC gives the tradeoff between Type I (false rejection) and Type II (false acceptance) error rates, denoted:
\begin{eqnarray}
\pI(g,\rO) & \equiv & \Pbb(g(\Dsf(\rO))=1 | \Hsf=0) \nonumber \\
\pII(g,\rO) & \equiv & \Pbb(g(\Dsf(\rO))=0 | \Hsf=1).
\end{eqnarray}
Recall that the null hypothesis $\Hsf=0$ corresponds to a failure of the reference transmission under the physical model.  A Type I error occurs for a realization of $\PPP$ such that the physical model fails ($\SINR_0 < \beta$) but the protocol model predicts success ($\|\xsf_i\| \geq \rO, \forall i \neq o$), i.e., the sum interference is ``large'' (enough to drive the SINR below the threshold) even though there are no ``near'' interferers.  A Type II error occurs for a realization of $\PPP$ such that the physical model succeeds ($\SINR_0 \geq \beta$) but the protocol model predicts failure ($\exists i \neq o : \|\xsf_i\| < \rO$), i.e., the sum interference is ``small'' even though there are one or more ``near'' interferers.

\begin{proposition}
\label{prp:errorprob}
The Type I and Type II error probabilities for $(g,\rO)$ are (with $\bar{g}(\cdot) \equiv 1 - g(\cdot)$):
\begin{eqnarray}
\label{eq:errorprob}
\pI(g) &=& \frac{\bar{p}_{\Hsf|\Dsf}(1|1) p_{\Dsf}(1)}{\bar{p}_{\Hsf}(1)}(g(1)-g(0)) + g(0) \nonumber \\
\pII(g) &=& \frac{p_{\Hsf|\Dsf}(1|1) p_{\Dsf}(1)}{p_{\Hsf}(1)}(\bar{g}(1)-\bar{g}(0)) + \bar{g}(0)
\end{eqnarray}
with $p_{\Hsf|\Dsf}(1|1), p_{\Hsf}(1), p_{\Dsf}(1)$ in terms of the model parameters in \prpref{posterior}, \lemref{prior}, and \lemref{evidence}. 
\end{proposition}

\begin{IEEEproof}
Apply standard probabilistic manipulations, including Bayes' rule.
\end{IEEEproof}

Under the uniform cost model the risk $R(\rO)$ \eqref{risk} reduces to the average error probability:
\begin{equation}
\label{eq:risk-uniform}
R(g,\rO) = \pI(g,\rO)\bar{p}_{\Hsf}(1) + \pII(g,\rO) p_{\Hsf}(1).
\end{equation}

\begin{figure}[ht]
\centering
\includegraphics[width=0.75\linewidth]{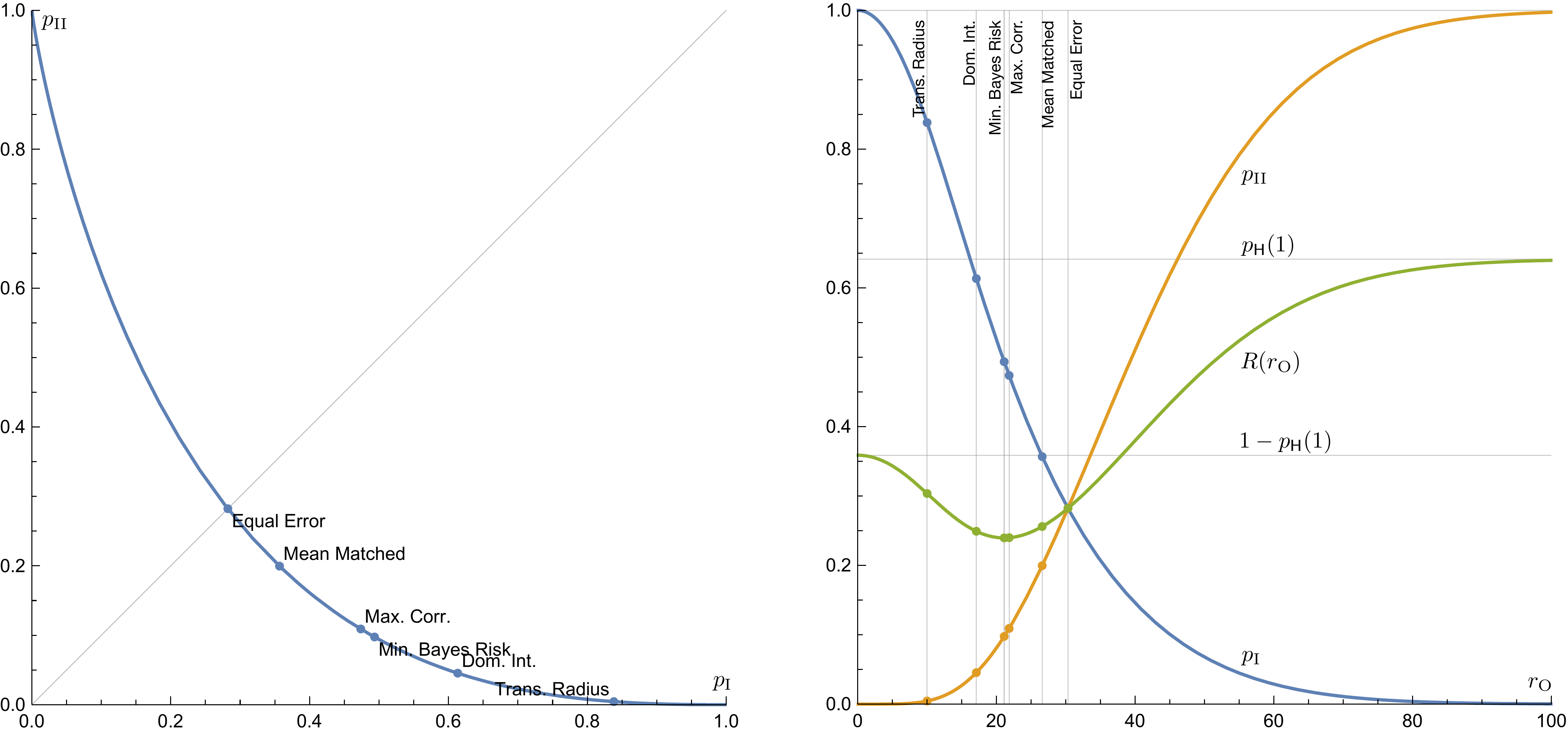}
\caption{ Uniform cost model, with parameters $n=2$, $\lambda = 2 \times 10^{-4}$, $\alpha = 3$, $\beta = 5$, $\rT = 10$, and $\eta = 0$.  
{\bf Left:} the receiver operating characteristic (ROC) of $(\pI,\pII)$ as $\rO$ is varied.  The equal error, mean matched, maximum correlation, minimum Bayes risk, dominant interferer, and transmission (i.e., $\rT$) radii are shown.
{\bf Right:} the Bayes risk $R(\rO)$, Type I error $\pI(\rO)$, and Type II error $\pII(\rO)$ vs.\ $\rO$.  
}
\label{fig:UCM}
\end{figure}

\begin{remark}
\label{rem:limiting-behavior}
Consider the rule $g(d) = d$, wherein protocol model success (failure) predicts physical model success (failure).  This rule behaves like the constant rules $g =0$ and $g=1$ in that $\Dsf = 1$ a.s.\ and $\Dsf = 0$ a.s., as $\rO \downarrow 0$ or $\rO \uparrow \infty$, respectively.  As $\rO \downarrow 0$, the protocol model will declare all transmissions successful (rejecting the null hypothesis with probability 1); however, the protocol model will \emph{falsely} reject $\Hsf=0$ with probability $p_{\Hsf}(0)$. As $\rO \uparrow \infty$, the protocol model will declare all transmissions fail (accepting the null hypothesis with probability 1); however, the protocol model will \emph{falsely} accept $\Hsf=1$ with probability $p_{\Hsf}(1)$. It follows that the asymptotic total risk $R(\rO)$ \eqref{risk-uniform} is $p_{\Hsf}(0)$ as $\rO \downarrow 0$ and $p_{\Hsf}(1)$ as $\rO \uparrow \infty$, i.e., the horizontal gridlines in \figref{UCM}.
\end{remark}

We now consider several specific operating points of the decision rule. First, interesting guard zone operating points include the extreme points as well as the TX-RX distance: $\rO \in \{0,\rT,\infty\}$. Second, we develop several additional guard zones in \lemref{gz-dominant-interferer}, \lemref{gz-mean-matched}, and \lemref{gz-equal-error}.

A {\em dominant interferer} (DI) under the physical model (without fading) is an interferer whose interference contribution is sufficient to violate the SINR threshold $\beta$.

\begin{lemma}
\label{lem:gz-dominant-interferer}
The minimum guard zone to exclude dominant interferers, is $\rODI \equiv \left(\frac{1}{\sigma} - \eta \right)^{-1/\alpha}$.
\end{lemma}

\begin{IEEEproof}
The $\rO$ solving $\frac{\rT^{-\alpha}}{\rO^{-\alpha} + \eta} = \beta$ prevents the existence of a dominant interferer.
\end{IEEEproof}

Define a guard zone $\rO$ to be {\em mean-matched} (MM) if the means of $\Dsf,\Hsf$ are equal, i.e., the probabilities of success under the physical and protocol models are equal.

\begin{lemma}
\label{lem:gz-mean-matched}
The mean-matched guard zone is $\rOMM \equiv \left(\kappa_{\delta} \sigma^\delta + \frac{\sigma\eta}{\lambda c_n} \right)^{1/n}$.
\end{lemma}

\begin{IEEEproof}
Set $p_{\Dsf}(1) = p_{\Hsf}(1)$ (i.e., $\Ebb[\Dsf] = \Ebb[\Hsf]$) and solve for $\rO$, via \lemref{prior} and \lemref{evidence}.
\end{IEEEproof}

Define an {\em equal error} (EE) guard zone as one with equal Type I and Type II error probabilities.

\begin{lemma}
\label{lem:gz-equal-error}
The guard zone achieving equal Type I and Type II errors, $\rOEE$, is the solution of:
\begin{equation}
1 = \erm^{-A} + \erm^{-B(\rO)} + \erm^{-C(\rO)} - 2\erm^{-A -C(\rO)}.
\end{equation}
\end{lemma}

\begin{IEEEproof}
Set $\pI(\rO) = \pII(\rO)$ and solve for $\rO$.
\end{IEEEproof}

\begin{remark}
\label{rem:gz-ordering}
We may readily obtain $\rODI \leq \rOMM$, since $\kappa_{\delta} \geq 1$. If $\beta \geq 1$, then we may further conclude that $\rT \leq \rODI \leq \rOMM$. 
\end{remark}

The operating points and error probabilities are illustrated in \figref{UCM}.  Observe that the minimum Bayes risk and maximum correlation radii are in close proximity for the chosen parameter values.  

\section{Impact of fading}
\label{sec:fading}

An asymmetry between the protocol and physical models is that success (failure) of the reference transmission under the {\em protocol} model, $\Dsf$, depends {\em solely} on the interfering TX locations $\{\xsf_i, i \in \Nbb\}$ from $\PPP$, while success (failure) under the {\em physical model}, $\Hsf$, depends on these locations {\em and} the (Rayleigh) random fades $\{\Fsf_i, i \in \Nbb\}$.  This raises the question: what fraction of the $i)$ loss in correlation between $(\Hsf,\Dsf)$ and $ii)$ Bayes risk (error) in estimating $\Hsf$ by observing $\Dsf$ is attributable to the fading in the physical model that is not captured in the protocol model?  

Besides Rayleigh fading, another special case for which some closed-form results are available is that of no fading ($\Fsf_i = 1$ for all $i$) and $\delta = 1/2$ (i.e., $\alpha = 2n$, which is $\alpha = 4$ for planar networks).  In this section we leverage these results to numerically investigate the impact of fading by comparing previous results for Rayleigh fading with new results for no fading.

Form the homogeneous marked PPP $\PPPb = \{(\xsf_i,\zsf_i), i \in \Nbb\}$ of intensity $\lambda > 0$, with $(\xsf_i,\zsf_i)$ as in $\PPP$.  That is, $\PPPb$ is $\PPP$ with the fades $\{\Fsf_i, i \in \Nbb\}$ removed.  Write $\tilde{\Hsf}$ for the success or failure of the reference transmission under the physical model with $\PPPb$.  Let $p_{\tilde{\Hsf}}(1) = \Pbb(\tilde{\Hsf}=1)$ denote the prior and $p_{\tilde{\Hsf}|\Dsf}(1|1) \equiv \Pbb(\tilde{\Hsf}=1|\Dsf=1)$ the posterior distribution.  Let $\tilde{\SINR}_o = \rT^{-\alpha}/(\tilde{\Isf}_o + \eta)$ be the corresponding SINR and $\tilde{\Isf}_o$ the sum interference at the reference RX. Define $Q(z)$ as the CCDF of the standard normal distribution $\Zsf \sim \Nmc(0,1)$. Observe 
\begin{equation}
p_{\tilde{\Hsf}}(1) = \Pbb(\tilde{\SINR}_o \geq \beta) = \Pbb(\tilde{\Isf}_o \leq 1/\sigma - \eta), 
\end{equation}
so the prior $p_{\tilde{\Hsf}}(1)$ is the CDF of $\tilde{\Isf}_o$ evaluated at $1/\sigma-\eta$.  The RV $\tilde{\Isf}_o$ has the L\'{e}vy distribution \cite[Definition 2.9]{WebAnd2012}, which has an ``explicit'' CDF in terms of $Q(\cdot)$.

\begin{proposition}
(\cite[Corollary 3.1]{WebAnd2012})
\label{prp:nofadeprior}
Fix $\delta = \frac{1}{2}$. The prior is
\begin{equation}
\label{eq:nofadeprior}
p_{\tilde{\Hsf}}(1) = 2Q \left( c_n \lambda \sqrt{\frac{\pi/2}{1/\sigma - \eta}}\right).
\end{equation}
\end{proposition}

\begin{proposition}
\label{prp:nofadelikelihood}
Fix $\delta = \frac{1}{2}$. The posterior may be obtained by numerically computing the ILT of the scaled LT for $\tilde{\Isf}_o$ (conditioned on the event $\Dsf = 1$), evaluated at $1/\sigma-\eta$:
\begin{equation}
\label{eq:nofadeCDFfromILT}
p_{\tilde{\Hsf}|\Dsf}(1|1) = \Lmc^{-1}\left\{ \frac{1}{s} \Lmc_{\tilde{\Isf}_o|\Dsf}(s|1) \right\}(1/\sigma - \eta).
\end{equation}
The LT of the sum interference $\tilde{\Isf}_o$, conditioned on the event $\Dsf = 1$, is
\begin{equation}
\Lmc_{\tilde{\Isf}_o|\Dsf}(s|1) = \exp \left(-\lambda c_n J(s,\rO^{-2n}) \right),
\end{equation}
for 
\begin{equation}
\label{eq:nofadltsumintjsu}
J(s,u) = \sqrt{\pi s} (1 - 2 Q(\sqrt{2 s u})) - \frac{1}{\sqrt{u}}(1 - \erm^{-s u}).
\end{equation}
\end{proposition}

\begin{figure}[ht]
\centering
\includegraphics[width=0.75\linewidth]{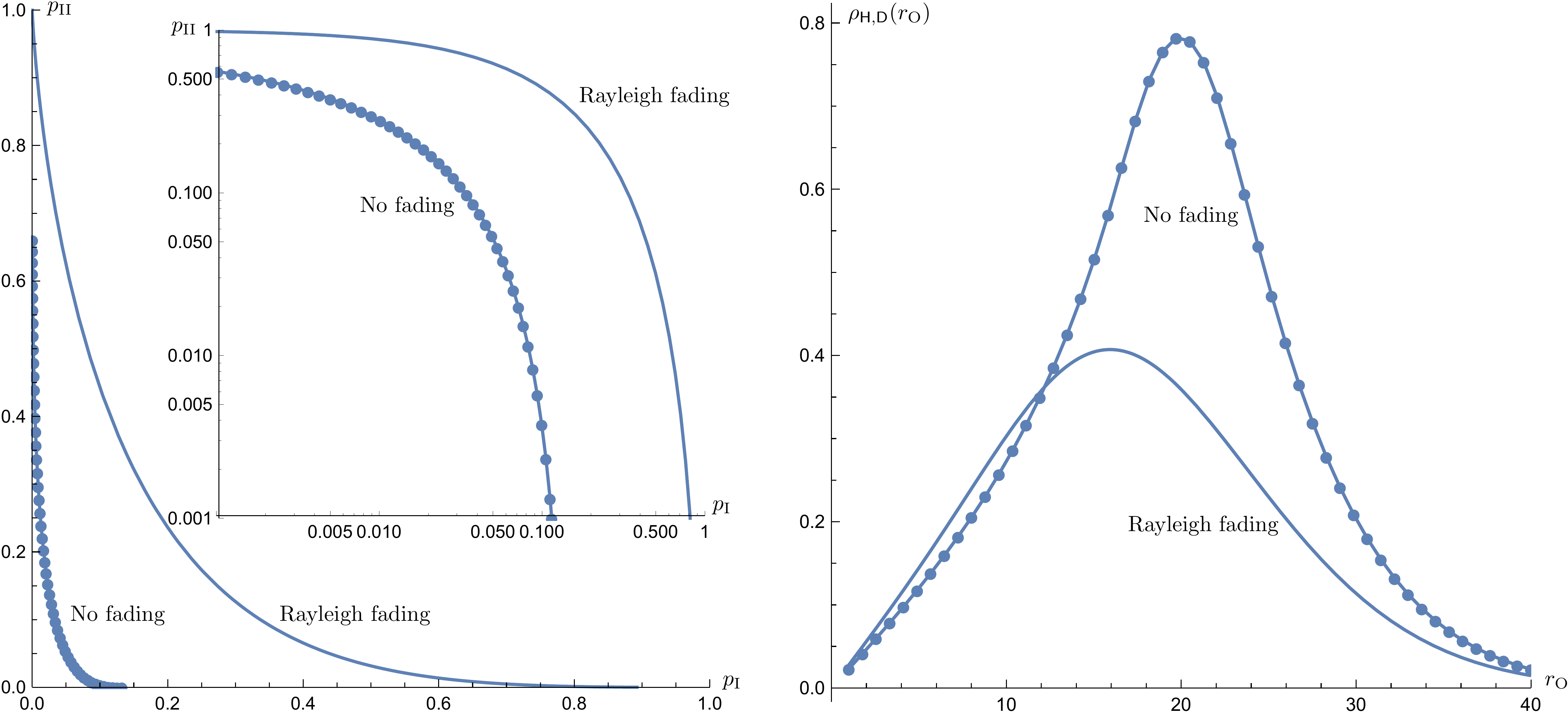}
\caption{
Impact of fading for $n=2$, $\lambda = 2 \times 10^{-3}$, $\alpha = 4$, $\beta = 5$, $\rT = 10$, $\eta = 0$.  
{\bf Left:} the ROC for the case of Rayleigh fading (solid line) and no fading (dotted line); the inset shows the same plot on logarithmic axes. 
{\bf Right:} the correlation $\rho_{\Hsf,\Dsf}(\rO)$ of $(\Hsf,\Dsf)$ vs.\ $\rO$ for Rayleigh fading (solid line) and no fading (dotted line).
}
\label{fig:nofading}
\end{figure}

The proof is in \prfref{nofadelikelihood}.  The ROC and the correlation with and without fading are shown in \figref{nofading}.  Both plots make clear that the absence of fading can (at least for the chosen parameter values) significantly increase (relative to Rayleigh fading) the utility of protocol model observations in inferring physical model success or failure.  In the ROC, for a wide range of values of $\pI$, we observe an order of magnitude (or more) improvement in $\pII$ (and vice-versa).  In the correlation plot we see a peak correlation of nearly $\approx 0.8$ without fading vs.\ $\approx 0.4$ with Rayleigh fading.  The ILT in \prpref{nofadelikelihood} was computed with \cite{ValAba2002}.

\section{Multiple protocol model observations under slotted Aloha}
\label{sec:multiobs}

Let time be slotted and indexed by $k \in \Nbb$. We next consider the case of multiple observations under the (slotted) Aloha protocol with parameter $p \in (0,1)$: each node attempts transmission at each time, independently of other nodes and across times, with probability $p$.  Let $\bar{p} \equiv 1-p$.  We assume throughout this section there is no noise, i.e., $\eta = 0$, and SINR reduces to SIR.  Let $\mathsf{\Phi}_{\rm {pot}} \equiv \{(\xsf_i,\zsf_i)\}$ be a homogeneous bipolar PPP of intensity $\lambda > 0$ representing the (random but fixed in time) locations of {\em potential} TX and RX, with TXs at $\{\xsf_i\}$ and RXs at $\{\ysf_i\}$, where $\ysf_i = \xsf_i + \zsf_i$.  Define the RVs $\Tsf \equiv (\Tsf_{i,k}, (i,k) \in \Nbb^2)$ with $\Tsf_{i,k} \sim \mathrm{Ber}(p)$, and $\Tsf_{i,k} = 1$ denoting that TX $i$ attempted transmission at time $k$.  Under Aloha, $\Tsf$ is IID across both nodes and times.  We further assume the time slot durations and fading coherence times are matched, with the idealization that the RVs $\Fsf \equiv (\Fsf_{i,k}, (i,k) \in \Nbb^2)$, with $\Fsf_{i,k}$ the random fade from TX $i$ to the reference receiver at $o$ at time $k$, are likewise IID across both nodes and times.  The process $\mathsf{\Phi}_{\rm {pot}}$ generates a sequence of identically distributed PPPs $(\mathsf{\Phi}_k, k \in \Nbb)$, with $\mathsf{\Phi}_k \subseteq \mathsf{\Phi}_{\rm {pot}}$ the PPP of attempted TX at time $k$, with intensity $\lambda p$, and $\Tsf_{i,k} = \mathbf{1}_{\xsf_i \in \mathsf{\Phi}_k}$.  Equivalently, we view $\mathsf{\Phi}_k = \{(\xsf_i,(\zsf_i,\Tsf_{i,k},\Fsf_{i,k})\}$ as the process $\mathsf{\Phi}_{\rm pot}$ augmented with IID marks $(\Tsf_{i,k},\Fsf_{i,k})$ for each $i \in \Nbb$.  The elements of $\{\mathsf{\Phi}_k\}$ are dependent due to their shared connection with $\mathsf{\Phi}_{\rm {pot}}$, but are conditionally independent given $\mathsf{\Phi}_{\rm {pot}}$, due to the independent transmission attempts and fades.  

Let $N \in \Nbb$ be the number of prior protocol model observations, in each of which the reference transmission has been attempted.  These observations produce a binary $N$-vector $d^{(N)} \equiv (d_1,\ldots,d_{N}) \in \{0,1\}^N$, where $d_k = 1$ ($0$) indicates that the reference transmission attempt was (not) successful under the protocol model at time $k$, for $k \in [N]$.  Observe $K_d^{(N)} \equiv \sum_{k \in [N]} d_k$, with $K_d^{(N)} \in \{0,\ldots,N\}$, is a sufficient statistic for $d^{(N)}$.  Given the $N$ observations $K_d^{(N)}$, for time $N+1$ the observer is given the knowledge of the outcome under the protocol model $d_{N+1} \in \{0,1\}$ and asked to predict the outcome under the physical model $h_{N+1} \in \{0,1\}$.  The corresponding RVs are $\Hsf_{N+1},\Dsf_{N+1},\Ksf_d^{(N)}$, but we henceforth in this section use the shorthand notation $\Hsf,\Dsf,\Ksf$.  We again use the Bayes risk framework, and restrict our attention to the uniform cost model \eqref{risk-uniform} from \secref{roc}.  We require the (prior) distribution of $\Hsf$, the (evidence) distribution of $\Dsf$, and the (posterior) distribution of $\Hsf$ given $\Dsf$, each conditioned on $\Ksf = K$.  

Let $\Gmc$ be the set of decision rules, where each rule $g \in \Gmc$ maps $(K,d) \in \{0,\ldots,N\} \times \{0,1\}$ to $h \in \{0,1\}$, with the interpretation $g$ predicts $\Hsf=h$ given inputs $(\Ksf,\Dsf) = (K,d)$.  There are $|\Gmc|=2^{(N+1)2}$ possible rules.  For each rule $g$ there is an associated partition of $\{0,\ldots,N\} \times \{0,1\}$ into two regions $(\Rmc_0(g),\Rmc_1(g))$, with $\Rmc_h(g) = \Rmc_{h,0}(g) \cup \Rmc_{h,1}(g)$, for $h \in \{0,1\}$, and into four subregions $(\Rmc_{h,d}(g), (h,d) \in \{0,1\}^2)$, with $\Rmc_{h,d}(g) = \{ (K,d) : g(K,d) = h\}$. These regions are used to compute the Type I and Type II error probabilities for each $g$:
\begin{eqnarray}
\pI(g) &\equiv& \Pbb((\Ksf,\Dsf) \in \Rmc_1(g) | \Hsf = 0) \nonumber \\
\pII(g) &\equiv& \Pbb((\Ksf,\Dsf) \in \Rmc_0(g) | \Hsf = 1).
\end{eqnarray}
Define notation: 
$p_{\Hsf}(h) \equiv \Pbb(\Hsf=h)$, 
$p_{\Hsf|\Dsf}(h|d) = \Pbb(\Hsf=h|\Dsf=d)$, 
$p_{\Hsf|\Ksf}(h|K) = \Pbb(\Hsf=h|\Ksf=K)$, 
$p_{\Hsf|\Ksf,\Dsf}(h|K,d) = \Pbb(\Hsf = h|\Ksf=K,\Dsf=d)$, 
$p_{\Ksf,\Dsf|\Hsf}(K,d|h) = \Pbb(\Ksf=K,\Dsf=d|\Hsf=h)$, 
$p_{\Dsf|\Ksf}(d|K) = \Pbb(\Dsf=d|\Ksf=K)$, and 
$p_{\Ksf}(K) = \Pbb(\Ksf=K)$. 
 \prpref{multiposterior} enables expression of the Type I, II error probabilities in \prpref{multiobserrorprobexp} in terms of computable quantities.

\begin{proposition}
\label{prp:multiposterior}
The (posterior) distribution $p_{\Hsf|\Ksf,\Dsf}(h|K,1)$ equals $p_{\Hsf|\Dsf}(h|1)$, i.e., the RVs $(\Hsf,\Ksf)$ are independent given $\Dsf=1$.  Moreover, $p_{\Hsf|\Dsf}(h|1)$ is given by \prpref{posterior} with $\lambda$ replaced by $p \lambda$.
\end{proposition}

\begin{IEEEproof}
The conditional independence holds since knowledge that $\Ksf = K$ (from which one can estimate $\Msf$, the number of potential TX in $b(o,\rO)$) has no bearing on $\Hsf$, given knowledge that $\Dsf=1$, i.e., none of the $\Msf$ {\em potential} TX in $b(o,\rO)$ transmit at time $N+1$.  The replacement of $\lambda$ by $p \lambda$ follows by the thinning property of the PPP.
\end{IEEEproof}

\begin{remark}
\label{rem:justificationmulti}
Although $(\Hsf,\Ksf)$ are conditionally independent given $\Dsf = 1$, they are {\em dependent} given $\Dsf = 0$.  Intuitively, knowledge of prior observations, summarized as $\Ksf$, {\em is} useful in estimating $\Hsf$ given $\Dsf = 0$, i.e., that one or more TX are active in $b(o,\rO)$.  A simple example shows this dependence.  Fix $\rO = 50$, $n=2$, $\lambda = 2 \times 10^{-4}$, $\alpha=3$, $\beta=5$, $\rT = 10$, $N=1$ and $p=1/2$, and compute $p_{\Hsf|\Dsf}(1|0) \approx 0.68$, $p_{\Hsf|\Ksf,\Dsf}(1|0,0) \approx 0.67$, $p_{\Hsf|\Ksf,\Dsf}(1|1,0) \approx 0.72$.  Thus, knowledge of protocol model success (failure) in the previous slot increases (decreases) the probability of physical model success in the current slot, given protocol model failure in the current slot.  This dependence justifies the study of decision rules $\Gmc$ with both $(K,d)$ as inputs in predicting $h$. 
\end{remark} 

\begin{proposition}
\label{prp:multiobserrorprobexp}
The Type I and Type II error probabilities under decision rule $g \in \Gmc$ are functions of $p_{\Hsf}(1),p_{\Hsf|\Dsf}(1|1),p_{\Hsf|\Ksf}(1|K),p_{\Dsf|\Ksf}(1|K),p_{\Ksf}(K)$ (with $\bar{p}(\cdot) \equiv 1-p(\cdot)$):
\begin{eqnarray}
\pI(g) &=& \frac{1}{\bar{p}_{\Hsf}(1)}\left( \bar{p}_{\Hsf|\Dsf}(1|1)  (\delta_{1,1}(g) - \delta_{1,0}(g)) + \delta_{\rm I}(g) \right) \nonumber \\
\pII(g) &=& \frac{1}{p_{\Hsf}(1)} \left( p_{\Hsf|\Dsf}(1|1) (\delta_{0,1}(g) - \delta_{0,0}(g)) +  \delta_{\rm II}(g) \right) \label{eq:multiobserrorprob}
\end{eqnarray}
where $\delta_{\rm I}(g) \equiv \sum_{K : (K,0) \in \Rmc_{1,0}(g)} \bar{p}_{\Hsf|\Ksf}(1|K) p_{\Ksf}(K)$, $\delta_{\rm II}(g) \equiv \sum_{K : (K,0) \in \Rmc_{0,0}(g)} p_{\Hsf|\Ksf}(1|K) p_{\Ksf}(K)$, and $\delta_{h,d}(g) \equiv \sum_{K : (K,d) \in \Rmc_{h,d}(g)} p_{\Dsf|\Ksf}(1|K) p_{\Ksf}(K)$.
\end{proposition}

\begin{IEEEproof}
Express $p_{\Ksf,\Dsf|\Hsf}(K,d|h)$ in $(\pI(g),\pII(g))$ in terms of $p_{\Hsf|\Ksf,\Dsf}(h|K,d)$ by Bayes' rule. \prpref{multiposterior} yields:
\begin{equation}
p_{\Hsf|\Ksf,\Dsf}(h|K,0) = \frac{p_{\Hsf|\Ksf}(h|K) - p_{\Hsf|\Dsf}(h|1)p_{\Dsf|\Ksf}(1|K)}{\bar{p}_{\Dsf|\Ksf}(1|K)}.
\end{equation}
\end{IEEEproof}

The prior $p_{\Hsf|\Ksf}(1|K)$ and evidence $p_{\Dsf|\Ksf}(1|K)$ distributions are given in \prpref{multipriorprot} and \prpref{multievidence}.  Define the Poisson RV $\Msf = \Msf(\mu_d)$, for $\mu_d \equiv \lambda c_n \rO^n$, as the number of potential inteferers inside the observation ball $b(o,\rO)$, i.e., $\Msf = \mathsf{\Phi}_{\rm pot}(b(o,\rO))$. \thmref{physucconmindisk} is of independent interest, but also is the key technical result required in the proof \prpref{multipriorprot}.

\begin{theorem}
\label{thm:physucconmindisk}
The distribution of the physical model feasibility RV $\Hsf$ given $\Msf = m$ potential interferers in $b(o,\rO)$, denoted $p_{\Hsf|\Msf}(1|m) \equiv \Pbb(\Hsf = 1 | \Msf =m)$, is (for $\xi \equiv \frac{p}{\bar{p}} \chi^{-\delta} I(\chi,\delta)$):
\begin{eqnarray}
\label{eq:multipriorgivprot}
p_{\Hsf|\Msf}(1|m) 
&=& \erm^{p \mu_d(1 - \chi^{-\delta}(\kappa_{\delta} + I(\chi,\delta)))} \times (1+\xi)^m \bar{p}^m.
\end{eqnarray}
\end{theorem}

The first (second) term is the probability the reference TX is successful under the physical model given interference from outside (inside) $b(o,\rO)$, when there are $m$ potential TX in $b(o,\rO)$.  

Define, for $k,l \in \Nbb$, $0 < a <1$, $\nu > 0$
\begin{equation}
\label{eq:multievidencefdfuncdef}
f_d(\nu,a;k,l) \equiv \sum_{j=0}^l \binom{l}{j} (-1)^j \erm^{-\nu(1- a^{k+j})}.
\end{equation} 

\begin{proposition}
\label{prp:multipriorprot}
The (prior) distribution of the physical model feasibility RV $\Hsf_{N+1}$ given $N$ protocol model observations $d^{(N)}$ with $K_d^{(N)}=K$ successes is (for $\xi \equiv \frac{p}{\bar{p}} \chi^{-\delta} I(\chi,\delta)$):
\begin{equation}
\label{eq:multiobsphK}
p_{\Hsf|\Ksf}(1|K) = \erm^{p \mu_d(1 - \chi^{-\delta} \kappa_{\delta} + \xi)} \frac{f_d(\mu_d(1+\xi),\bar{p},K+1,N-K)}{f_d(\mu_d,\bar{p},K,N-K)}.
\end{equation}
\end{proposition}

\begin{proposition}
\label{prp:multievidence}
The (evidence) distribution of the protocol model feasibility RV $\Dsf_{N+1}$ given $N$ protocol model observations $d^{(N)}$ with $K_d^{(N)}=K$ successes is (for $\mu_d \equiv \lambda c_n d_{\rm I}^n$ and $f_d$ in \eqref{multievidencefdfuncdef}):
\begin{equation}
\label{eq:multiprior}
p_{\Dsf|\Ksf}(1|K) = \frac{f_d(\mu_d,\bar{p};K+1,N-K)}{f_d(\mu_d,\bar{p};K,N-K)}.
\end{equation}
\end{proposition}

\begin{figure}[ht]
\centering
\includegraphics[width=0.75\linewidth]{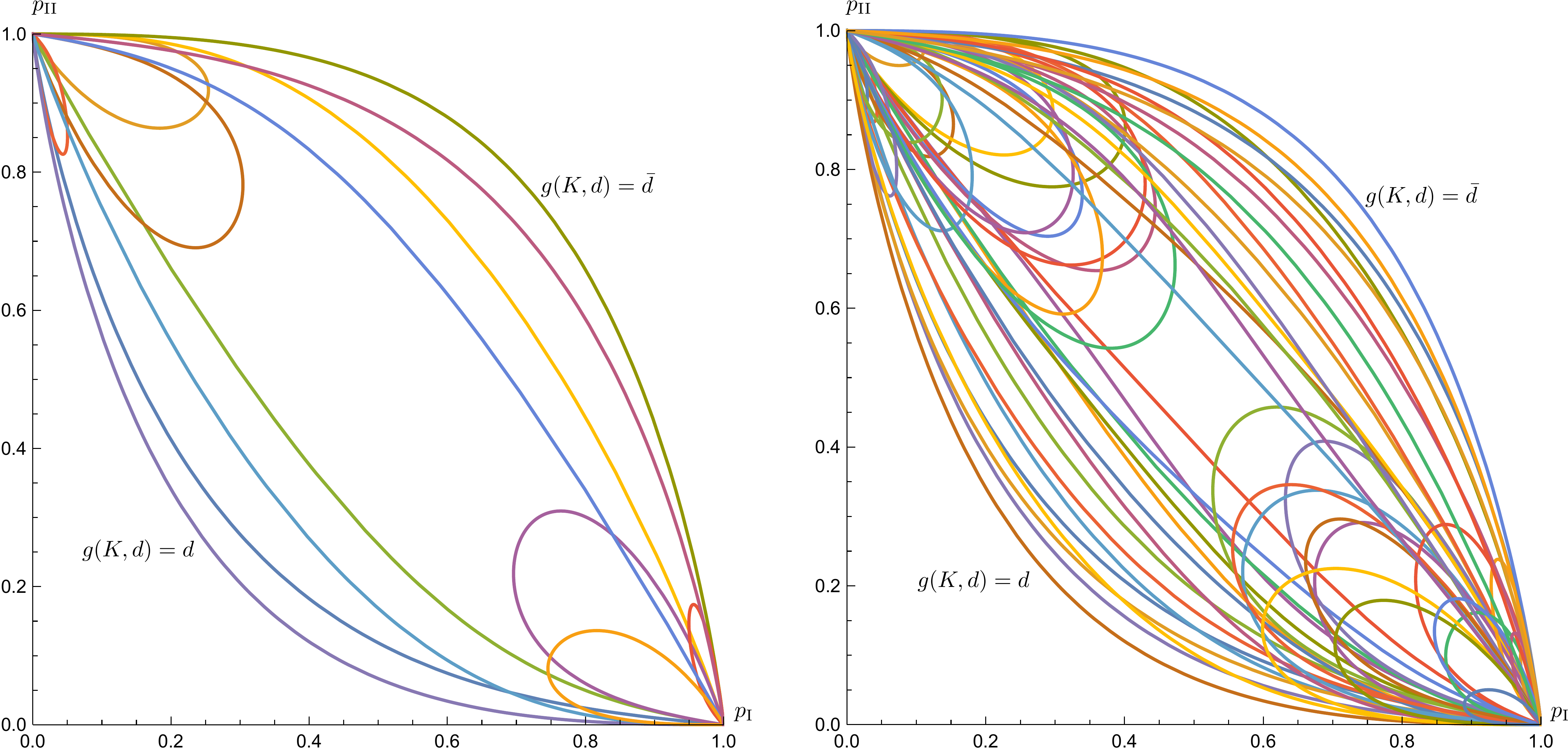}
\caption{
ROC for each of the $|\Gmc| = 2^{(N+1)2}$ possible decision rules under $N$ prior protocol model observations, for $n=2$, $\lambda = 2 \times 10^{-4}$, $\alpha = 3$, $\beta = 5$, $\rT = 10$, $\eta = 0$, with $N=1$ (left) and $N=2$ (right).  In both cases the best decision rule is $g(K,d) = d$ and the worst is $g(K,d) = \bar{d}$.
}
\label{fig:multipleobs}
\end{figure}

Proofs of \thmref{physucconmindisk}, \prpref{multipriorprot}, \prpref{multievidence} are given in \prfref{physucconmindisk}, \prfref{multipriorprot}, \prfref{multievidence} respectively.  The ROCs for all possible decision rules, for $N \in \{1,2\}$ observations, are shown in in \figref{multipleobs}.  In both cases the optimal decision rule is $g(K,d) = d$, i.e., to ignore the prior observations (despite the correlation of $(\Ksf,\Dsf)$) and simply guess $\Hsf = d$.  

\section{Conclusions}
\label{sec:conclusions}

Our five sets of results (\secref{corr} through \secref{multiobs}) have analyzed the connection between the protocol and physical interference models.  With so many papers in wireless communications and networking written for one (but not the other) of these two models, our primary contribution is to have partially illuminated the probabilistic connection between them.  The suggestion from \figref{multipleobs} that previous protocol model observations may not be useful in the optimal decision rule given the current protocol model outcome, despite their dependence with physical model success (\remref{justificationmulti}), motivates our ongoing investigations into the role played by previous protocol and/or physical model observations in predicting future protocol and/or physical model success.  

\appendix

\subsection{Proof of \prpref{posterior}}
\label{prf:posterior}

\begin{IEEEproof}
The homogeneous PPP $\PPP$, conditioned on the event $\Dsf = 1$, is stochastically equivalent to a nonhomogeneous PPP $\PPP_{\rO}$ with a radially isotropic intensity function $\lambda_{\rO}(x)$, with parameters $\rO \geq 0$ and $\lambda > 0$, that excludes TXs within distance $\rO$ from the origin $o$: 
\begin{equation}
\label{eq:pppintensvoidball}
\lambda_{\rO}(x) \equiv \lambda \mathbf{1}\left\{\|x\| \geq \rO \right\}, ~ x \in \Rbb^d.
\end{equation}
Write $\SINR_o(\PPP)$ and $\SINR_o(\PPP_{\rO})$ for the SINR and $\Isf_o(\PPP)$ and $\Isf_o(\PPP_{\rO})$ for the sum interference at the reference receiver under these two processes.  Then:
\begin{eqnarray}
p_{\Hsf|\Dsf}(1|1)
&=& \Pbb(\SINR_o(\PPP) \geq \beta | \Dsf = 1) \nonumber \\
&=& \Pbb(\SINR_o(\PPP_{\rO}) \geq \beta) \nonumber \\
&\overset{(a)}{=}& \Pbb(\Fsf_o \geq \beta \rT^{\alpha} \Isf_o(\PPP_{\rO}) ) \Pbb(\Fsf_o \geq \beta \rT^{\alpha} \eta)  \nonumber \\
&\overset{(b)}{=}& \Lmc_{\Isf_o|\Dsf}(\sigma|1) \erm^{-\sigma\eta}.
\end{eqnarray}
In \emph{(a)} we expand $\SINR_o$, isolate $\Fsf_o$, and apply the memoryless property of $\Fsf_o$.  In \emph{(b)} we recognize the first term is the LT of $\Isf_o(\PPP_{\rO})$ from PPP $\PPP_{\rO}$; the second term is the CCDF of $\Fsf_o$ with $\sigma \equiv \beta\rT^{\alpha}$. Finally, we employ \corref{laplace-with-void} below to evaluate the LT $\Lmc_{\Isf_o|\Dsf}(s|1)$ of $\Isf_o(\PPP_{\rO})$ with transmitter-free void-zone radius $\rO$, i.e., conditioned on $\Dsf = 1$, at $s = \sigma$.
\end{IEEEproof}

\subsection{Laplace transform (LT) of sum interference over a PPP with void ball}
\label{app:laplace-with-void}

The LT of the sum interference observed at the origin $\Isf_o \equiv \sum_{i \neq o} \Fsf_i l(\|\xsf_i\|)$ under the PPP $\PPP_{\rO}$, i.e., conditioned on $\Dsf = 1$, is denoted $\Lmc_{\Isf_o|\Dsf}(s|1) \equiv \Ebb_{\PPP_{\rO}}[\erm^{-s\Isf_o}]$.

\begin{corollary}(of \cite[p.103]{Hae2013})
\label{cor:laplace-with-void}
Let $\PPP_{\rO}$ be a marked PPP on $\Rbb^n$ with isotropic intensity function $\lambda_{\rO}(x)$ \eqref{pppintensvoidball}.  The LT of the sum interference $\Isf_o$ observed at the origin is (for $I(u,\delta)$ in \eqref{convfcn}):
\begin{equation}
\label{eq:LTaggIntdI}
\log \Lmc_{\Isf_o|\Dsf}(s|1) = 
\lambda c_n( \rO^n - \kappa_{\delta} s^{\delta} - s^{\delta} I(\rO^{\alpha}/s,\delta)).
\end{equation}
\end{corollary}

\begin{IEEEproof}
Straightforward adaptation of the development in \cite[p.103]{Hae2013} to our scenario yields:
\begin{equation}
\log \Lmc_{\Isf_o|\Dsf}(s|1) 
= - \lambda c_n \Ebb \left[ n \int_{\rO}^{\infty}  \left(1 - \erm^{-s \Fsf r^{-\alpha}}\right) r^{n-1} \drm r \right], \label{eq:laplace1}
\end{equation}
The integral (with $q = s \Fsf$) may be expressed in terms of $E(v,u)$ and $\Gamma(v)$ in \eqref{stdfcndef}:
\begin{equation}
n \int_{\rO}^{\infty}  \left(1 - \erm^{-q r^{-\alpha}}\right) r^{n-1} \drm r 
= -\rO^n + q^{\delta} \Gamma(1-\delta) + \rO^n\delta E(1+\delta,q \rO^{-\alpha}).
\end{equation}
Substitution of $s \Fsf$ for $q$ and linearity of expectation gives
\begin{eqnarray}
\log \Lmc_{\Isf_o|\Dsf}(s|1) 
&=& -\lambda c_n \Ebb \left[ -\rO^n + (s \Fsf)^{\delta} \Gamma(1-\delta) + \rO^n\delta E(1+\delta,s \Fsf \rO^{-\alpha}) \right] \nonumber \\
&=& \lambda c_n (\rO^n - s^{\delta} \Gamma(1-\delta)\Ebb[\Fsf^{\delta}] - \rO^n\delta \Ebb[E(1+\delta,s \Fsf \rO^{-\alpha})]) 
\end{eqnarray}
Observe $\Ebb[\Fsf^{\delta}] = \Gamma(1+\delta)$ for $\Fsf$ a unit rate exponential $\Fsf$.  Recalling the LT of $\Fsf$ is $\Lmc_{\Fsf}(s) = 1/(1+s)$, and using the change of variables $q' = s \rO^{-\alpha}$ and $t' = 1/(q't)$ allows:
\begin{eqnarray}
\Ebb[E(1+\delta,q' \Fsf )]
&=& \Ebb \left[ \int_{1}^{\infty} \erm^{-q't \Fsf} t^{-(1+\delta)} \drm t \right] \nonumber \\
&=& \int_{1}^{\infty} \Ebb \left[ \erm^{-q't \Fsf} \right] t^{-(1+\delta)} \drm t  \nonumber \\
&=& \int_{1}^{\infty} \Lmc_{\Fsf}(q't) \; t^{-(1+\delta)} \drm t  \nonumber \\
&=& \int_{1}^{\infty} \frac{1}{1+q't} t^{-(1+\delta)} \drm t  \nonumber \\
&=& \frac{(q')^{\delta}}{\delta} \delta \int_0^{1/q'} \frac{(t')^{\delta}}{1+t'} \drm t' 
\end{eqnarray}
Substitution gives
\begin{equation}
\log \Lmc_{\Isf_o|\Dsf}(s|1) 
= \lambda c_n \left(\rO^n - s^{\delta} \Gamma(1-\delta) \Gamma(1+\delta) - \rO^n\delta \frac{(q')^{\delta}}{\delta} I(1/q',\delta) \right) 
\end{equation}
Substituting in $q'$ and using the definitions of $\kappa_{\delta}$ and $\delta$ gives \eqref{LTaggIntdI}.
\end{IEEEproof}

\subsection{Proof of \prpref{coroptobs}}
\label{prf:coroptobs}

\begin{IEEEproof}
We show the four properties in turn. 
$i)$ $\lim_{\chi \downarrow 0} \rho(\chi) = 0$.  Observe $B(0) = C(0) = 0$. Substituting $\chi = 0$ gives the indeterminate form $\rho(0) = 0/0$.  L'Hopital's rule, using $C'(\chi) = B'(\chi) \chi/(1+\chi)$ and $B'(\chi) = \delta B(\chi)/\chi$, gives:
\begin{equation}
\lim_{\chi \downarrow 0} \rho(\chi) = \lim_{\chi \downarrow 0} \frac{2 \sqrt{\erm^{B(\chi)}-1}}{(1+\chi) \erm^{C(\chi)}} = \frac{0}{1} = 0.
\end{equation}

$ii)$ $\lim_{\chi \uparrow \infty} \rho(\chi) = 0$.  Observe
\begin{equation}
\label{eq:chidelminIchidel}
\chi^{\delta} - I(\chi,\delta)
= \delta \int_0^{\chi} t^{\delta-1} \drm t - \delta \int_0^{\chi} \frac{t^{\delta}}{1+t} \drm t = \delta \int_0^{\chi} \frac{t^{\delta-1}}{1+t} \drm t \geq 0
\end{equation}
and, recalling $\kappa_{\delta}$ from \eqref{kappadef}, we see 
\begin{equation}
\chi^{\delta} - I(\chi,\delta) \leq \delta \int_0^{\infty} \frac{t^{\delta-1}}{1+t} \drm t = \kappa_{\delta}.
\end{equation}
It follows that $0 < B(\chi) - C(\chi) \leq \lambda c_n \sigma^{\delta} \kappa_{\delta}$.  Therefore, as $\lim_{\chi \uparrow \infty} B(\chi) = \infty$, 
\begin{equation}
\lim_{\chi \uparrow \infty} \rho(\chi) \leq \lim_{\chi \uparrow \infty} \frac{\erm^{\lambda c_n \sigma^{\delta} \kappa_{\delta}}-1}{\sqrt{\erm^{B(\chi)}-1}} = 0. 
\end{equation}
Given $\rho (\chi) > 0$ for all $\chi > 0$ (property $iii)$ below), it follows that $\lim_{\chi \uparrow \infty} \rho(\chi) = 0$.

$iii)$  $\rho(\chi) \in (0,1]$ for all $\chi > 0$.  Observe $B(\chi) > C(\chi)$ implies $\rho(\chi) > 0$ in \eqref{corHDexp}, and $B(\chi) > C(\chi)$ is equivalent to $\chi^{\delta} > I(\chi,\delta)$, shown in \eqref{chidelminIchidel}.

$iv)$  has a maximum at $\chi^* > 1$ equal to a positive solution of  \eqref{coroptobs}.  The first derivative, simplified using $B'(\chi),C'(\chi)$ from $i)$:
\begin{equation}
\rho'(\chi)= \rho(\chi) B'(\chi) \left[ \frac{1}{(1-\erm^{-(B(\chi)-C(\chi))})(1+\chi)}-\frac{1}{2(1-\erm^{-B(\chi)})} \right]. 
\end{equation}
That $\chi^*$ is an extremum follows by observing $\rho'(\chi) = 0$ is equivalent to either $\chi = 0$ or the expression in square brackets being equal to zero, which may be rearranged as \eqref{coroptobs}.  Should multiple solutions to \eqref{coroptobs} exist, at least one of them must correspond to the global maximum, by virtue of the fact that $\rho(0)=\rho(\infty) = 0$ and $\rho(\chi) > 0$.  We now establish the existence of a solution to \eqref{coroptobs}.  Observe \eqref{coroptobs} may be equivalently written as $f_1(\chi) = f_2(\chi)$ for $f_1(\chi) \equiv (1-\chi) \erm^{B(\chi)}$ and $f_2(\chi) \equiv 2 - (1+\chi)\erm^{C(\chi)}$.  The argument is to first establish $a)$ $f_1(\chi) - f_2(\chi) > 0$ over $\chi \in (0,1)$ (hence no solutions in $(0,1)$), and to then prove $b)$ there exists a solution to $f_1(\chi) = f_2(\chi)$ over $\chi \in [1,\infty)$.  We first prove $a)$.  
\begin{eqnarray}
f_1(\chi) - f_2(\chi) > 0 
& \Leftrightarrow & (1-\chi)\erm^{B(\chi)} + (1+\chi)\erm^{C(\chi)} > 2 \nonumber \\
& \Leftrightarrow & (1-\chi)\erm^{B(\chi)-C(\chi)} + (1+\chi) > 2 \erm^{-C(\chi)}.
\end{eqnarray}
By item $ii)$ above, $B(\chi) - C(\chi) > 0$, ensuring $(1-\chi)\erm^{B(\chi)-C(\chi)} + (1+\chi) > 2$ for $\chi \in (0,1)$, which, when combined with $C(\chi) \geq 0$, proves the statement.  We now prove $b)$.  First observe the function $g_2(\chi) \equiv \log((1+\chi)/(1-\chi))$ for $\chi > 1$ has derivatives $g_2'(\chi) = -2/(\chi^2-1) < 0$ and $g_2^{''}(\chi) = 4 \chi/(\chi^2-1)^2 > 0$, and as such obeys $i)$ $\lim_{\chi \downarrow 1} g_2(\chi) = \infty$, $ii)$ $\lim_{\chi \uparrow \infty} g_2(\chi) = 0$, $iii)$ $g_2(\chi) > 0$ for $\chi > 1$, and $iv)$ $g_2(\chi)$ is convex decreasing over $\chi > 1$.  Next observe the equation $(1-\chi) \erm^{B(\chi)} + (1+\chi) \erm^{C(\chi)} = 0$ is equivalent to $g_1(\chi) = g_2(\chi)$ (for $g_1(\chi) \equiv B(\chi)-C(\chi)$) and to $f_1(\chi) = f_2(\chi)-2$.  From \eqref{chidelminIchidel}, we know $g_1(\chi)$ is increasing in $\chi > 0$ with $g_1(0) = 0$ and $\lim_{\chi \uparrow \infty} g_1(\chi) = \kappa_{\delta}$.  It follows there exists $\hat{\chi} > 1$ such that $g_1(\hat{\chi}) = g_2(\hat{\chi})$, and therefore $f_1(\hat{\chi}) = f_2(\hat{\chi}) - 2$, and in particular $f_1(\hat{\chi}) < f_2(\hat{\chi})$.  At $\chi = 1$ we have $f_2(1) = 2(1-\erm^{C(2)}) < 0 = f_1(1)$.  By the fixed point theorem, as the continuous functions $(f_1(\chi),f_2(\chi))$ obey $f_1(1) > f_2(1)$ and $f_1(\hat{\chi}) < f_2(\hat{\chi})$, there must exist $\chi^* \in [1,\hat{\chi}]$ at which $f_1(\chi) = f_2(\chi)$.
\end{IEEEproof}

\subsection{Proof of \thmref{min-risk}}
\label{prf:min-risk}

\begin{IEEEproof}
Recall $R(\rO)$ and $(A,B(\rO),C(\rO))$ are defined in \eqref{risk2}, \prpref{posterior}, respectively.  For conciseness, we will refer to $R(\rO)$, $B(\rO)$ and $C(\rO)$ without arguments, and write $\gamma \equiv c_{10}-c_{00}$ and $\nu \equiv c_{01}-c_{11}$.  All derivatives are with respect to $\rO$.  The first two derivatives of $R$ are:
\begin{eqnarray}
R' &=& - \gamma B' \erm^{-B} + (\nu + \gamma) C' \erm^{-A-C} \nonumber \\
R'' &=& \gamma \left((B')^2- B''\right) \erm^{-B} - (\nu + \gamma) \left((C')^2 - C''\right) \erm^{-A-C}, \label{eq:rd2}
\end{eqnarray}
with: $B' = n \lambda c_n \rO^{n - 1}$, $B''= \frac{n - 1}{\rO} B'$, 
$C' = B' \frac{\chi}{1 +\chi}$, and $C'' = \frac{(n - 1)(1 + \chi) + \alpha}{\rO (1 + \chi)} C'$.
Assume henceforth that $c_{10} > c_{00}$ ($\gamma > 0$) and $c_{01} > c_{11}$ ($\nu > 0$), ensuring $\nu/\gamma > 0$.  We establish conditions for existence and uniqueness and then prove quasi-convexity. 

{\em Existence and uniqueness.}
The equation $R' = 0$ may be rearranged as
\begin{equation}
\label{eq:rd1-zero}
\frac{B'}{C'} = \left(1 + \frac{\nu}{\gamma} \right) \exp(-A+B-C),
\end{equation}
which is equivalent to \eqref{gz-min-risk}, and then rearranged into the form $f_L(\rO) = f_R(\rO)$, where 
\begin{eqnarray}
f_L & \equiv & \log \left(1+\frac{1}{\chi} \right) - \log \left(1+\frac{\nu}{\gamma} \right) \nonumber \\
f_R & \equiv & - \lambda c_n \kappa_{\delta} \sigma^{\delta} - \sigma \eta + \lambda c_n \rO^n - \lambda c_n \sigma^{\delta} I(\chi,\delta).
\end{eqnarray}
These functions have derivatives $f_L' =  \frac{-\alpha}{\rO \sigma(1+\chi)} < 0$ and $f_R' =\frac{n \lambda c_n \rO^{n-1}}{1+\chi} > 0$, 
and limiting values 
$\lim_{\rO \downarrow 0} f_L(\rO) = \infty$, 
$\lim_{\rO \uparrow \infty} f_L(\rO) = - \log(1 + \nu/\gamma) < 0$, 
$\lim_{\rO \downarrow 0} f_R(\rO) = -\lambda c_n \kappa_{\delta} \sigma^{\delta} - \sigma \eta < 0$, and
$\lim_{\rO \uparrow \infty} f_R(\rO) = - \sigma \eta < 0$.
Of these, the only one of any difficulty is $\lim_{\rO \uparrow \infty} f_R(\rO)$.  To prove the limit, write $\Delta_R \equiv \lim_{\rO \uparrow \infty} f_R(\rO) - f_R(0)$ and use the change of variable $\chi = \rO^{\alpha}/\sigma$:
\begin{eqnarray}
\Delta_R 
&=& \lambda c_n \sigma^{\delta} \lim_{\chi  \uparrow \infty}  \left( \chi^{\delta} - I(\chi ,\delta) \right) \nonumber \\
&=& \lambda c_n \sigma^{\delta} \lim_{\chi  \uparrow \infty}  \left( \delta \int_0^{\chi} t^{\delta-1} \drm t - \delta \int_0^{\chi} \frac{t^{\delta}}{1+t}\drm t \right) \nonumber \\
&=& \lambda c_n \sigma^{\delta} \delta \int_0^{\infty} \frac{t^{\delta-1}}{1+t}\drm t = \lambda c_n \sigma^{\delta} \kappa_{\delta}
\end{eqnarray}
using \eqref{kappadef}. In summary, $f_L(\rO)$ is decreasing from $+\infty$ down to $- \log(1 + \nu/\gamma) < 0$, while $f_R(\rO)$ is increasing from $-\lambda c_n \kappa_{\delta} \sigma^{\delta} - \sigma \eta < 0$ up to $-\sigma \eta < 0$.  It is clear an intersection of $f_L(\rO),f_R(\rO)$ will exist iff $-\log(1+\nu/\gamma) < - \sigma\eta$, yielding \eqref{existtest}.  If the condition holds, it follows by the monotonicity of the two functions that the intersection is unique. 

{\em Quasi-convexity.} We employ a sufficient condition for quasi-convexity \cite[Eq.~3.22]{BoyVan2004}:
\begin{equation}
\label{eq:quasiconvex}
R'(\rO) = 0 \implies R''(\rO) > 0, \quad \forall \rO \in (0,\infty).
\end{equation}
To establish the sufficient condition holds, evaluate \eqref{rd2} at a stationary point $\rO^*$ obeying \eqref{rd1-zero}:
\begin{eqnarray}
R''({\rO^*})
&=& \gamma \left((B')^2- B''\right) \erm^{-B} - \gamma \frac{B'}{C'} \left((C')^2 - C''\right) \erm^{-B} \\
&=& \gamma B' \erm^{-B} \left( \left(B' - \frac{B''}{B'}\right) - \left(C' - \frac{C''}{C'} \right) \right) \\
&=& \gamma B' \erm^{-B} \left( (B' - C') + \left(\frac{C''}{C'} - \frac{B''}{B'}\right) \right) > 0
\end{eqnarray}
where $B' > 0$, $\erm^{-B} > 0$, $B' - C' = B'\left(1 - \frac{\chi}{1 + \chi}\right) > 0$, and $\frac{C''}{C'} - \frac{B''}{B'} = \frac{\alpha}{\rO(1+\chi)} > 0$. 
Thus $R(\rO)$ is quasi-convex in $\rO$.  \figref{BayesRisk} shows $R(\rO)$ is not in general convex.  Finally, \eqref{optrisk} follows by substituting \eqref{gz-min-risk} into \eqref{bayes-risk-ppp}.
\end{IEEEproof}

\subsection{Proof of \prpref{sensitivity}}
\label{prf:sensitivity}

\begin{IEEEproof}
Let $\zeta$ denote either parameter $(\lambda,\sigma)$ to be studied.  Recall the definitions of $\nu,\gamma$ from \prfref{min-risk}, the change of variable $\chi = \chi(\sigma,\rO) = \rO^{\alpha}/\sigma$, and define $D(\zeta,\rO) \equiv \log(1 + 1/\chi)$ (observing $\chi$ depends on both $\lambda,\sigma$).  Then $R'(\rO)=0$ in \eqref{gz-min-risk} may be written (with $(A,B(\rO),C(\rO))$ in \prpref{posterior}) as $g(\zeta,\rO) = 0$, with:
\begin{equation}
g(\zeta,\rO) = D(\zeta,\rO) - \log(1+\nu/\gamma) + A(\zeta) - B(\zeta,\rO) + C(\zeta,\rO). \label{eq:gpgen}
\end{equation}
By \thmref{min-risk}, $\rO^*$ is the unique solution of $g(\zeta,\rO^*) = 0$.  By the implicit function theorem, the sensitivity of $\rO^*$ to parameter $\zeta$ is
\begin{equation}
\label{eq:dIoptder}
\frac{\drm \rO^*}{\drm \zeta} = - \frac{\frac{\partial}{\partial \zeta} g(\zeta,\rO)}{\frac{\partial}{\partial \rO} g(\zeta,\rO)}.
\end{equation}
We require the following partial derivatives (recall $\eta = 0$, by assumption), presented in ``Jacobian form'' for functions $\{A,B,C,D\}$ and arguments $\{\rO,\lambda,\sigma\}$:
\begin{equation}
\label{eq:parder}
\kbordermatrix{
     & \rO & \lambda & \sigma \\
A &   & c_n \kappa_{\delta} \sigma^{\delta}  & \lambda c_n \kappa_{\delta} \delta \sigma^{\delta-1} \\
B & \lambda c_n n \rO^{n-1} & c_n \rO^{n} & \\
C & \lambda c_n n \rO^{n-1}\frac{\chi}{1+\chi} & c_n \sigma^{\delta} I(\chi,\delta) & \lambda c_n \delta\left( \sigma^{\delta-1}I(\chi,\delta) - \frac{\rO^n \chi}{\sigma(1+\chi)}\right)  \\
D & -\frac{\alpha}{\rO(1+\chi)} & & \frac{1}{\sigma(1+\chi)}  
}
\end{equation}
The three empty entries indicate the function is independent of the parameter or variable.  

{\em Sensitivity of $\rO^*$ to $\lambda$.} Substitution and algebra, using \eqref{gpgen}, \eqref{dIoptder}, and \eqref{parder}, yields \eqref{dIstdlam}.
To show $\frac{\drm }{\drm \lambda} \rO^* > 0$, it suffices to show $\Delta_{\lambda}$, defined below, is positive (recall \eqref{kappadef}):
\begin{eqnarray}
\Delta_{\lambda} 
&\equiv & \kappa_{\delta} + I(\chi,\delta) - \chi^{\delta} \nonumber \\
&=& \delta \int_0^{\infty} \frac{t^{\delta-1}}{1+t} \drm t + \delta \int_0^{\chi} \frac{t^{\delta}}{1+t} \drm t - \delta \int_0^{\chi} t^{\delta-1} \drm t \nonumber \\
& = & \delta \int_{\chi}^{\infty} \frac{t^{\delta-1}}{1+t} \drm t > 0 \label{eq:deltalambdaineq}
\end{eqnarray}

{\em Sensitivity of $\rO^*$ to $\sigma$.} Substitution and algebra, using \eqref{gpgen}, \eqref{dIoptder}, and \eqref{parder}, yields \eqref{dIstds}.
To show $\frac{\drm }{\drm \sigma} \rO^* > 0$, it suffices to show $\Delta_{\sigma} \equiv (1+\chi)(\kappa_{\delta} + I(\chi,\delta)) - \chi^{\delta+1}$ is positive.  To show $\Delta_{\sigma} > 0$, view $\Delta_{\sigma}(\chi)$ as a function of $\chi$ on $\Rbb_+$ and prove both $\Delta_{\sigma}(0) > 0$ and $\Delta_{\sigma}'(\chi) > 0$; together this ensures $\Delta_{\sigma} > 0$.  First, $\Delta_{\sigma}(0) = \kappa_{\delta} > 0$.  Second, for $\Delta_{\lambda}$ in \eqref{deltalambdaineq},
\begin{eqnarray}
\Delta_{\sigma}'(\chi) &=& (\kappa_{\delta} + I(\chi,\delta)) + (1+\chi) \delta \frac{\chi^{\delta}}{1+\chi} - (\delta+1) \chi^{\delta} \nonumber \\
&=& \kappa_{\delta} + I(\chi,\delta) - \chi^{\delta} = \Delta_{\lambda} > 0. 
\end{eqnarray}
\end{IEEEproof}

\subsection{Proof of \prpref{nofadelikelihood}}
\label{prf:nofadelikelihood}

\begin{IEEEproof}
Recall the use of $\PPP_{\rO}$ to create the transmitter-free null-zone for realizations of $\PPP$ consistent with the conditioned event $\Dsf = 1$ in \prfref{posterior}.  Analogously, we define the nonhomogeneous PPP $\PPPb_{\rO} = \{(\xsf_i,\zsf_i), i \in \Nbb\}$ with a radially isotropic intensity function $\lambda_{\rO}(x)$ in \eqref{pppintensvoidball} to achieve the same effect for $\PPPb$.  The likelihood function 
\begin{eqnarray}
p_{\tilde{\Hsf}|\Dsf}(1|1) 
&=& \Pbb(\tilde{\SINR}_o(\PPPb) \geq \beta |\Dsf=1) \nonumber \\
&=& \Pbb(\tilde{\SINR}_o(\PPPb_{\rO}) \geq \beta) \nonumber \\
&=& \Pbb(\tilde{\Isf}_o(\PPPb_{\rO}) \leq 1/{\sigma}-\eta), \label{eq:nofadelikelihoodpf1}
\end{eqnarray}
is thus the CDF of the sum interference seen at the reference receiver $\tilde{\Isf}_o$ under $\PPPb_{\rO}$ evaluated at $t = 1/{\sigma}-\eta$.  Write $F_{\tilde{\Isf}_o|\Dsf}(t|1)$, and $\Lmc_{\tilde{\Isf}_o|\Dsf}(s|1)$ for the CDF and LT of $\tilde{\Isf}_o$ under $\PPPb_{\rO}$.  As evident from \eqref{nofadelikelihoodpf1}, the likelihood requires the CDF of $\tilde{\Isf}_o$.  Although it is not available explicitly for any $\rO > 0$, (it {\em is} available explicitly for the case $\rO = 0$, as in \prpref{nofadeprior}), we can obtain it by {\em numerically} computing the inverse LT via the basic identity (c.f.\ \eqref{nofadeCDFfromILT}):
\begin{equation}
F_{\tilde{\Isf}_o|\Dsf}(t|1) = \Lmc^{-1}\left\{ \frac{1}{s} \Lmc_{\tilde{\Isf}_o|\Dsf}(s,1) \right\}(t).
\end{equation}

The truncated power law impulse response (pathloss) function $l_{\alpha,\epsilon}(r) \equiv r^{-\alpha} \mathbf{1}_{r \geq \epsilon}$ \cite[Eq.\ 2.22]{WebAnd2012} has a null-zone of radius $\epsilon$ around the receiver.  Observe the equivalence between $i)$ the sum interference seen at the origin $\tilde{\Isf}_o$ under the nonhomogeneous PPP $\PPPb_{\rO}$ with (non-truncated) pathloss $l(r) \equiv r^{-\alpha}$, and $ii)$ the sum interference $\tilde{\Isf}_o$ under the homogeneous PPP $\PPPb$ with truncated pathloss $l_{\alpha,\epsilon}$, provided we set $\epsilon = \rO$.  The LT of the latter is provided in \cite[Corollary 2.5 (c.f.\ Eq. 2.51)]{WebAnd2012}:
\begin{equation}
\Lmc_{\tilde{\Isf}_o|\Dsf}(s|1) = \exp \left( - \lambda c_n \delta \int_0^{\rO^{-\alpha}} (1-\erm^{-s y}) y^{-\delta-1} \drm y \right).
\end{equation}
The result follows by specializing to $\delta = 1/2$, and integrating
\begin{equation}
J(s,u) \equiv \frac{1}{2} \int_0^u (1-\erm^{-s y}) y^{-3/2} \drm y.
\end{equation}  
to obtain \eqref{nofadltsumintjsu}.  The tractability for $\rO = 0$ ($u = \infty$) for the prior \prpref{nofadeprior} is due to $J(s,\infty) = \sqrt{\pi s}$.
\end{IEEEproof}

\subsection{Proof of \thmref{physucconmindisk}}
\label{prf:physucconmindisk}

\begin{IEEEproof}
Let RVs $\Hsf,\Fsf_o,\Isf_o$ be the physical model success indicator, reference signal fade, and interference seen at $o$, all for time $N+1$.  Condition on $\Msf$ given in the theorem:
\begin{eqnarray}
p_{\Hsf|\Msf}(1|m)
&=& \Pbb(\Fsf_o \geq \sigma \Isf_o |\Msf=m) \nonumber \\
&=& \Ebb[\Pbb(\Fsf_o \geq \sigma \Isf_o |\Isf_o,\Msf=m)|\Msf=m] \nonumber \\
&=& \Ebb[\erm^{- \sigma \Isf_o} |\Msf=m].
\end{eqnarray}
Define point processes $(\tilde{\mathsf{\Phi}}, \tilde{\mathsf{\Phi}}_{\rm in}, \tilde{\mathsf{\Phi}}_{\rm out})$ with: $i)$ $\tilde{\mathsf{\Phi}} = \tilde{\mathsf{\Phi}}_{\rm in} \cup \tilde{\mathsf{\Phi}}_{\rm out}$, $ii)$ $\tilde{\mathsf{\Phi}}_{\rm in} = (\xsf_1,\ldots,\xsf_m)$, with IID uniform RVs $\xsf_i \sim \mathrm{Uni}(b(o,\rO))$ for $i \in [m]$, and $iii)$ $\tilde{\mathsf{\Phi}}_{\rm out} = (\xsf_{m+1},\xsf_{m+2},\ldots)$ a PPP with radially isotropic intensity function $\lambda_{\rO}(x)$ \eqref{pppintensvoidball}.  By construction, $\tilde{\mathsf{\Phi}}$ is a PPP of intensity $\lambda$ outside of $b(o,\rO)$ and with $m$ points uniformly distributed over $b(o,\rO)$, and the points give the positions of  potential TX.  Observe $(\tilde{\mathsf{\Phi}}_{\rm in}, \tilde{\mathsf{\Phi}}_{\rm out})$ are independent.  Define $(\tilde{\Isf},\tilde{\Isf}_{\rm in},\tilde{\Isf}_{\rm out})$ as the interference seen at $o$ at time $N+1$ generated by the three point processes above using the general form:
\begin{equation}
\Isf = \sum_{i \in \mathsf{\Phi}} \Tsf_i \Fsf_i \|\xsf_i\|^{-\alpha}.
\end{equation}
where $\Tsf_i$ is the contention decision of TX $i$ and $\Fsf_i$ is the fade from TX $i$ to the reference RX, both at time $N+1$.  Observe $\tilde{\Isf}=\tilde{\Isf}_{\rm in}+\tilde{\Isf}_{\rm out}$ and $(\tilde{\Isf}_{\rm in},\tilde{\Isf}_{\rm out})$ are independent.  By construction
\begin{equation}
\Ebb[\erm^{- \sigma \Isf_o} |\Msf=m] 
= \Lmc_{\tilde{\Isf}}(\sigma) 
= \Lmc_{\tilde{\Isf}_{\rm in}}(\sigma) \Lmc_{\tilde{\Isf}_{\rm out}}(\sigma).
\end{equation}  
It remains to find the two LTs $\Lmc_{\tilde{\Isf}_{\rm in}}(s)$ and $\Lmc_{\tilde{\Isf}_{\rm out}}(s)$. 

\underline{LT of $\tilde{\Isf}_{\rm in}$.}  The independence of the locations of the $m$ points in $\tilde{\mathsf{\Phi}}_{\rm in}$ allows:
\begin{equation}
\Lmc_{\tilde{\Isf}_{\rm in}}(s)
= \Ebb \left[ \prod_{i \in [m]} \erm^{-s \Tsf_i \Fsf_i \|\xsf_i\|^{-\alpha}} \right]
= \Ebb \left[ \erm^{-s \Tsf \Fsf \|\xsf\|^{-\alpha}} \right]^m 
\end{equation}
where $(\Tsf,\Fsf)$ here denote an arbitrary member $(\Tsf_{i,k},\Fsf_{i,k})$.  
Recalling $\Tsf \sim \mathrm{Ber}(p)$:
\begin{equation}
\Ebb \left[ \erm^{-s \Tsf \Fsf \|\xsf\|^{-\alpha}} \right]
= \Ebb \left[ \erm^{-s \Fsf \|\xsf\|^{-\alpha}} \right] p + \bar{p} 
\end{equation}
Conditioning on $\xsf$, using the LT of the exponential distribution,  recalling $\xsf \sim \mathrm{Uni}(b(o,\rO))$, leveraging the radial symmetry of the pathloss function to transform the integral from $n$ down to $1$ dimensions, using the change of variables $r' = r^{\alpha}/s$, and recalling $I(u,\delta)$ in \eqref{convfcn} gives:
\begin{eqnarray}
\Ebb \left[ \erm^{-s \Fsf \|\xsf\|^{-\alpha}} \right]
&=& \Ebb[ \Ebb[\erm^{-s \Fsf \|\xsf\|^{-\alpha}}|\xsf]] \nonumber \\
&=& \Ebb \left[ \frac{1}{1+s\|\xsf\|^{-\alpha}} \right] \nonumber \\
&=& \frac{1}{c_n \rO^n} \int_{b(o,\rO)} \frac{1}{1+s\|x\|^{-\alpha}} \drm x \nonumber \\
&=& \frac{n}{\rO^n}  \int_0^{\rO} \frac{1}{1+s r^{-\alpha}} r^{n-1} \drm r \nonumber \\
&=& \rO^{-n} s^{\delta} \delta \int_0^{\rO^{\alpha}/s} \frac{(r')^{\delta}}{1+r'} \drm r' 
\end{eqnarray}
Substitution and the parameters $\chi \equiv \rO^{\alpha}/\sigma$ and $\xi \equiv \frac{p}{\bar{p}} \chi^{-\delta} I(\chi,\delta)$ gives $\Lmc_{\tilde{\Isf}_{\rm in}}(\sigma) = (1+\xi)^m \bar{p}^m$.

\underline{LT of $\tilde{\Isf}_{\rm out}$.}  Recall \prpref{posterior}, the LT of the interference seen at $o$ conditioned on there being no points in the observation ball $b(o,\rO)$, was derived for the single observation setting with transmitter intensity $\lambda$.  By the thinning property of the PPP, it applies here with intensity $p \lambda$:
\begin{eqnarray}
\label{eq:LTaggIntdI2}
\log \Lmc_{\tilde{\Isf}_{\rm out}}(\sigma) 
&=& p \lambda c_n( \rO^n - \kappa_{\delta} \sigma^{\delta} - \sigma^{\delta} I(\chi,\delta)) \nonumber \\
&=& p \mu_d ( 1 - \chi^{-\delta}(\kappa_{\delta}  +  I(\chi,\delta)))
\end{eqnarray}
\end{IEEEproof}

\subsection{Proof of \prpref{multipriorprot}}
\label{prf:multipriorprot}

Denote by $\mathrm{Po}(\nu)$ the Poisson distribution with parameter $\nu > 0$, and by $\mathrm{Po}(m;\nu)$ its PMF evaluated at $m \in \Zbb_+$.  We will have cause to use the following three lemmas.

\begin{lemma}
\label{lem:poissonexpect}
For $0 < a < 1$, $\nu > 0$, let $\Msf_1,\Msf_2$ be Poisson RVs with PMFs $\mathrm{Po}(m;\nu),\mathrm{Po}(m;a\nu)$, respectively.  Then $\Ebb[a^{\Msf_1}g(\Msf_1)] = \erm^{-\nu(1-a)} \Ebb[g(\Msf_2)]$ for any measurable function $g : \Nbb \to \Rbb_+$.
\end{lemma}

\begin{IEEEproof}
By definition of expectation, the two sides of the equation below prove the result:
\begin{equation}
\sum_{m=0}^{\infty} a^m g(m) \mathrm{Po}(m;\nu) = \erm^{-\nu(1-a)} \sum_{m=0}^{\infty} g(m) \mathrm{Po}(m;a \nu). 
\end{equation}
\end{IEEEproof}

Define, for $k,l,m \in \Nbb$ and $0<a<1$
\begin{equation}
g_d(m,a;k,l) \equiv (a^m)^k(1-a^m)^l.
\end{equation}

\begin{lemma}
\label{lem:poissonmoment}
Let $M \sim \mathrm{Po}(\nu)$, $a \in (0,1)$, and $k,l \in \Nbb$.  Then, for $f_d$ in \eqref{multievidencefdfuncdef}:
\begin{equation}
\Ebb[g_d(\Msf,a;k,l)] = \Ebb[(a^{\Msf})^k(1-a^{\Msf})^l] = f_d(\nu,a;k,l).
\end{equation}
\end{lemma}

\begin{IEEEproof}
Apply the binomial theorem, use linearity of expectation, and apply \lemref{poissonexpect}:
\begin{equation}
\Ebb[g_d(\Msf,a;k,l)] = \sum_{j=0}^l \binom{l}{j} (-1)^j \Ebb[(a^{k+j})^{\Msf}] = \sum_{j=0}^l \binom{l}{j} (-1)^j \erm^{-\nu(1-a^{k+j})}.   
\end{equation}
\end{IEEEproof}

Recall the RV $\Msf \sim \mathrm{Po}(\mu_d)$, for $\mu_d \equiv \lambda c_n \rO^n$ is the number of points from $\mathsf{\Phi}_{\rm pot}$ in $b(o,\rO)$.  It is used in the following lemma and the proof of \prpref{multipriorprot}.

\begin{lemma}
\label{lem:poissoncondprotobs}
The probability of there being $\Msf = m$ points from $\mathsf{\Phi}_{\rm pot}$ in $b(o,\rO)$, given $K$ successes out of $N$ protocol model observations, $\Pbb(\Msf = m | \Ksf = K)$, is:
\begin{equation}
\label{eq:multiobsprooffirstgoal}
\frac{\Pbb(\Msf = m | \Ksf = K)}{\Pbb(\Msf = m)} = \frac{g_d(m,\bar{p};K,N-K)}{f_d(\mu_d,\bar{p};K,N-K)}.
\end{equation}
\end{lemma}

\begin{IEEEproof}
By Bayes' rule:
\begin{equation}
\label{eq:bayesruleform}
\frac{\Pbb(\Msf = m | \Ksf = K)}{\Pbb(\Msf = m)}
= \frac{\Pbb( \Ksf = K|\Msf = m)}{\Pbb(\Ksf = K)}.
\end{equation}
The numerator is the binomial PMF with $K$ successes in $N$ trials with success probability $\bar{p}^m$:
\begin{equation}
\label{eq:numeratorexp}
\Pbb( \Ksf = K|\Msf = m) = \binom{N}{K} g_d(m,\bar{p};K,N-K).
\end{equation}
Trial $k$ is successful, meaning $d_k = 1$, when none of the $m$ TX from $\mathsf{\Phi}_{\mathrm {pot}}$ in $b(o,\rO)$ transmit, which happens with probability $\bar{p}^m$.  The denominator is found by conditioning on $\Msf$ and \lemref{poissonmoment}:
\begin{eqnarray}
\Pbb(\Ksf = K) 
&=& \Ebb[ \Pbb(\Ksf = K|\Msf)] \nonumber \\
&=& \Ebb \left[ \binom{N}{K} g_d(\Msf,\bar{p};K,N-K) \right] \nonumber \\
&=& \binom{N}{K} \Ebb \left[ g_d(\Msf,\bar{p};K,N-K) \right] \label{eq:multiobsuncondprobkofnsuc}
\end{eqnarray}
Applying \lemref{poissonmoment} to \eqref{multiobsuncondprobkofnsuc} and substituting it and \eqref{numeratorexp} into \eqref{bayesruleform} yields \eqref{multiobsprooffirstgoal}.
\end{IEEEproof}

\begin{IEEEproof}[Proof of \prpref{multipriorprot}]
Condition on $\Msf$, and use the independence of $(\Hsf,\Ksf)$ given $\Msf$: 
\begin{eqnarray}
p_{\Hsf|\Ksf}(1|K) &=& \Ebb[\Pbb(\Hsf=1|\Msf, \Ksf = K)|\Ksf = K] \nonumber \\
&=& \Ebb[\Pbb(\Hsf=1|\Msf)|\Ksf = K] \nonumber \\
&=& \sum_{m=0}^{\infty} \Pbb(\Hsf=1|\Msf=m)\Pbb(\Msf=m|\Ksf=K)
\end{eqnarray}
Use $\Pbb(\Msf=m|\Ksf=K)$ from \lemref{poissoncondprotobs} and $\Pbb(\Hsf=1|\Msf=m)$ from \thmref{physucconmindisk}, yielding $p_{\Hsf|\Ksf}(1|K) = p_{h,{\rm out}}(K) p_{h,{\rm in}}(K)$ with (recall $\xi \equiv \frac{p}{\bar{p}} \chi^{-\delta} I(\chi,\delta)$):
\begin{eqnarray}
p_{h,{\rm out}}(K) &=& \frac{\erm^{p \mu_d ( 1 - \chi^{-\delta}(\kappa_{\delta}  +  I(\chi,\delta)))}}{f_d(\mu_d,\bar{p},K,N-K)} \nonumber \\
p_{h,{\rm in}}(K) &=& \Ebb[ (1+\xi)^{\Msf} \bar{p}^{\Msf} g_d(\Msf,\bar{p},K,N-K)]
\end{eqnarray}
Use $a^m g_d(m,a;k,l) = g_d(m,a;k+1,l)$, then use \lemref{poissonexpect} with $\Msf_1 = \Msf \sim \mathrm{Po}(\mu_d)$ and $\Msf_2 \sim \mathrm{Po}(\mu_d(1+\xi))$,  and finally use \lemref{poissonmoment}:
\begin{eqnarray}
p_{h,{\rm in}}(K) 
&=& \Ebb\left[ \left(1+\xi \right)^{\Msf_1} g_d(\Msf_1,\bar{p},K+1,N-K) \right] \nonumber \\
&=& \erm^{\mu_d \xi} \Ebb\left[ g_d(\Msf_2,\bar{p},K+1,N-K) \right] \nonumber \\
&=& \erm^{\mu_d \xi} f_d(\mu_d(1+\xi),\bar{p},K+1,N-K) 
\end{eqnarray}
Lastly, combine and simplify the two exponents to obtain \eqref{multiobsphK}.
\end{IEEEproof}

\subsection{Proof of \prpref{multievidence}}
\label{prf:multievidence}

\begin{IEEEproof}
Recall from \prfref{multipriorprot} that $\Msf \equiv \mathsf{\Phi}_{\rm {pot}}(b(o,\rO)) \sim \mathrm{Po}(\mu_d)$ is the number of points from $\mathsf{\Phi}_{\mathrm {pot}}$ in the observation ball $b(o,\rO)$. It is clear that $\Msf$ is a sufficient statistic for estimating $\Dsf$ from $\mathsf{\Phi}_{\mathrm {pot}}$.  Condition on $\Msf$, use the independence of $(\Dsf,\Ksf)$ given $\Msf$, use $\Pbb(\Dsf=1|\Msf=m)=\bar{p}^m$, use $\Pbb(\Msf = m | \Ksf = K)$ from \lemref{poissoncondprotobs}, and use $a^m g_d(m,a;k,l) = g_d(m,a;k+1,l)$:
\begin{eqnarray}
p_{\Dsf|\Ksf}(1|K)
&=& \Ebb\left[ \Pbb( \Dsf=1|\Msf,\Ksf= K) |\Ksf= K \right] \nonumber \\ 
&=& \Ebb\left[ \Pbb( \Dsf=1|\Msf) |\Ksf= K \right] 
= \sum_{m=0}^{\infty} \bar{p}^m \Pbb(\Msf = m | \Ksf= K) \nonumber \\
&=& \sum_{m=0}^{\infty} \bar{p}^m \frac{g_d(m,\bar{p};K,N-K)}{f_d(\mu_d,\bar{p};K,N-K)} \mathrm{Po}(m;\mu_d) \nonumber \\
&=& \frac{\Ebb \left[ g_d({\Msf},\bar{p};K+1,N-K) \right]}{f_d(\mu_d,\bar{p};K,N-K)} 
\end{eqnarray}
Applying \lemref{poissonmoment} to the numerator proves the result.
\end{IEEEproof}

\vspace{-0.2in}

\bibliographystyle{IEEEtran}
\bibliography{2016-08-TWC-JW}

\end{document}